# Survey of AI-Powered Approaches for Osteoporosis Diagnosis in Medical Imaging

Abdul Rahman[a] and Bumshik Lee[b, *]

[a] *Department of Information and Communication Engineering, Chosun University, Gwangju 61452, Republic of Korea*

[b] *Energy AI, Korea Institute of Energy Technology (KENTECH), Naju 58330, South Korea*

[*] *Corresponding author, bslee@kentech.ac.kr*

**Abstract**

Osteoporosis silently erodes skeletal integrity worldwide; however, early detection through imaging can prevent most fragility fractures. Artificial intelligence (AI) methods now mine routine Dual-energy X-ray Absorptiometry (DXA), X-ray, Computed Tomography (CT), and Magnetic Resonance Imaging (MRI) scans for subtle, clinically actionable markers, but the literature is fragmented. This survey unifies the field through a tri-axial framework that couples imaging modalities with clinical tasks and AI methodologies (classical machine learning, convolutional neural networks (CNNs), transformers, self-supervised learning, and explainable AI). Following a concise clinical and technical primer, we detail our Preferred Reporting Items for Systematic Reviews and Meta-Analyses (PRISMA)-guided search strategy, introduce the taxonomy via a roadmap figure, and synthesize cross-study insights on data scarcity, external validation, and interpretability. By identifying emerging trends, open challenges, and actionable research directions, this review provides AI scientists, medical imaging researchers, and musculoskeletal clinicians with a clear compass to accelerate rigorous, patient-centered innovation in osteoporosis care. The project page of this survey can also be found on [Github](Github).

## 1. Introduction

Osteoporosis, often referred to as a *silent disease,* is a systemic metabolic bone disorder that develops without clinical symptoms until a fracture occurs. Characterized by a reduction in bone mineral density (BMD) and deterioration of bone microarchitecture, osteoporosis compromises the structural integrity of the skeleton, thereby increasing susceptibility to fragility fractures. These fractures, particularly those of the hip, spine, and wrist, can result in significant pain, disability, loss of independence, and even mortality, especially among the elderly population. This condition is most prevalent in postmenopausal women due to hormonal changes; however, it also affects men and younger individuals with secondary risk factors, such as chronic diseases, long-term medication use, and malnutrition.

Osteoporosis is a growing global health concern, affecting an estimated 200 million people worldwide. In the United States alone, over 10 million individuals are diagnosed with osteoporosis, and an additional 44 million have low bone mass, placing them at increased risk of developing the disease [1]. As the global life expectancy increases, the incidence of osteoporosis-related fractures is projected to increase sharply, exerting a substantial burden on healthcare systems in terms of hospitalizations, surgical interventions, long-term care, and lost productivity. Therefore, from a public health perspective, the early detection and prevention of osteoporotic fractures are of paramount importance.

Despite its serious consequences, osteoporosis remains significantly underdiagnosed and undertreated. This is primarily due to its asymptomatic nature in the early stages, as well as limitations in current diagnostic workflows. Dual-energy X-ray absorptiometry (DXA) is widely regarded as the clinical gold standard for assessing BMD and is used to define osteoporosis based on the World Health Organization's (WHO's) T-score thresholds. However, DXA alone often fails to capture the full spectrum of fracture risks. To improve risk stratification, the Fracture Risk Assessment Tool (FRAX) [2], which integrates clinical risk factors with or without BMD input, was developed to estimate an individual's 10-year probability of hip or major osteoporotic fractures. DXA and FRAX are now used complementarily in many clinical guidelines to support therapeutic decision-making and enhance preventive care strategies [3]. Nevertheless, studies have shown that up to 50% of postmenopausal women who experience vertebral fractures may present with normal BMD when assessed using DXA, and only a small proportion exhibit both radiographically confirmed fractures and DXA-defined osteoporosis [4]. These findings underscore the need for more advanced, predictive, and multidimensional diagnostic approaches that extend beyond conventional measurements.

This diagnostic gap is further exacerbated by the inherent ill-posedness of osteoporosis classification (i.e., lack of stability and completeness in the diagnostic criteria), which is currently based on a rigid threshold (T-score ≤ –2.5) that oversimplifies the continuous and multifactorial nature of fracture risk. Patient-specific factors such as age, sex, hormonal status, comorbid conditions, medication history, and lifestyle are not explicitly captured in imaging-derived BMD scores. Moreover, variability in imaging quality, overlapping morphological features, such as cortical thinning and trabecular pattern loss, and subclinical vertebral deformities further obscure accurate diagnosis. These complexities require diagnostic solutions that can analyze multimodal data, uncover latent patterns, and integrate diverse clinical variables into a unified framework. In this context, artificial intelligence (AI) has emerged as a transformative tool capable of modeling complex, high-dimensional relationships and extracting subtle cues from medical images that are often imperceptible to human observers.

AI has already demonstrated remarkable success across a variety of clinical imaging domains. In radiology, convolutional neural networks (CNNs) have been employed to detect lung nodules in chest computed tomography (CT) scans [5], identify breast lesions in mammography [6], and localize abnormalities in brain magnetic resonance imaging (MRI) with expert-level accuracy [7]. In ophthalmology, AI algorithms have been deployed in real-world screening programs for diabetic retinopathy [8] and age-related macular degeneration using retinal fundus images [9]. Similarly, in dental diagnostics, recent studies have leveraged deep learning (DL) models to achieve highly accurate localization of individual teeth and pathology, despite anatomical overlaps and noise [10]. The strength of DL lies in its ability to learn hierarchical feature representations directly from raw medical images, eliminating the need for handcrafted features and enabling significant performance gains across various imaging modalities and diagnostic tasks [11]. These advancements underscore the versatility of AI in interpreting complex anatomical structures and predicting disease risk across heterogeneous data sources. The integration of AI with osteoporosis-related imaging techniques, such as DXA, X-ray, and CT, holds similar potential, not only to automating BMD estimation and fracture detection but also to stratify patients based on nuanced risk profiles can potentially uncover previously unrecognized biomarkers of bone fragility.

This survey presents a structured and in-depth review of AI-powered osteoporosis detection techniques, with a primary focus on medical imaging modalities such as DXA, X-ray, CT, and MRI. Previous surveys have explored isolated aspects that are often constrained by modality, algorithmic scope, or specific clinical tasks. Our survey addresses these limitations by offering a modality- and task-aware synthesis that bridges clinical relevance and technical rigor. The main contributions of this survey are as follows:

- We systematically review AI methods applied across four major imaging modalities–DXA, X-ray, CT, and MRI–highlighting both modality-specific innovations and cross-modality comparisons.
- We classify the literature by diagnostic tasks, such as osteoporosis classification, fracture detection, and fracture risk prediction, offering a more practical and application-oriented perspective.
- We present an in-depth analysis of AI techniques, ranging from conventional machine learning (ML) and CNNs to emerging paradigms such as the Transformer and self-supervised learning. Special attention is given to model evaluation practices, training strategies, and interpretability.
- We analyze trends across time and modalities, identify methodological and translational gaps, and highlight underexplored combinations of imaging techniques and AI approaches.
- The included studies are examined for dataset availability, external validation, clinical integration potential, and explainability to facilitate an evidence-based assessment of real-world applicability.

By focusing on clinical relevance, technical depth, and a structured taxonomy, this survey aims to provide a comprehensive reference for researchers, healthcare professionals, and developers working on AI-based approaches for the early detection and management of osteoporosis.

## 2. Related Surveys

With the increasing prevalence of osteoporosis and asymptomatic progression, AI-driven imaging has emerged as a promising frontier in early detection and risk assessment. Several surveys have attempted to map this evolving landscape, focusing on ML and DL applications for bone health evaluation. Early works primarily reviewed classical ML approaches using DXA-derived BMD for osteoporosis classification. More recent efforts have shifted towards DL, particularly CNNs, applied to modalities such as X-ray and CT for tasks like vertebral fracture detection and T-score prediction. However, these surveys are often narrow in scope and limited to a single imaging modality, specific diagnostic tasks, or a selection of AI techniques. Thus, they fail to capture the diversity of available methods and the broader clinical context.

To better understand the strengths and limitations of existing reviews, we summarize them in **Table 1**. While some surveys have attempted a broader cross-modality synthesis, such as those covering CT, X-ray, DXA, and MRI, they often lack algorithmic details, imaging-specific methodologies, or structured task-wise analyses. Others provide strong clinical relevance or focused modality insights but fall short on technical depth, standardization, or comparative evaluation. Notably, few surveys have addressed newer learning paradigms, such as the Transformer or self-supervised models, and discussions on explainability, reproducibility, and real-world integration are often sparse.

**Table 1.** Summary of existing surveys

| Ref. | Year | Dataset Description | Imaging Modalities | | | | Remarks |
|---|---|---|---|---|---|---|---|
| | | | X-ray | CT | DXA | MRI | |
| [12] | 2021 | ✗ | ✓ | ✓ | ✓ | ✗ | A comprehensive survey of ML applications in osteoporosis across various modalities, but it lacks depth in imaging-specific methodologies and algorithmic detail. |
| [13] | 2021 | ✗ | ✓ | ✓ | ✗ | ✗ | While this survey offers a broad overview of ML applications in osteoporosis across various clinical and diagnostic modalities, it falls short in delivering detailed insights into imaging-specific techniques and the underlying algorithmic frameworks. |
| [14] | 2021 | ✗ | ✗ | ✗ | ✗ | ✓ | This review focuses on trabecular bone analysis using MRI and ML; however, it lacks comparative insights and comprehensive model evaluations. |
| [15] | 2023 | ✗ | ✗ | ✓ | ✗ | ✗ | This review highlights AI-driven osteoporosis classification using CT images but remains limited in task coverage and the range of imaging modalities considered. |
| [16] | 2023 | ✗ | ✗ | ✗ | ✓ | ✗ | This meta-analysis provides detailed insights into ML-based diagnosis using hip DXA data but is constrained by limited imaging variety and the breadth of ML applications. |
| [17] | 2024 | ✗ | ✓ | ✗ | ✗ | ✗ | This modality-focused review demonstrates strong clinical relevance and model diversity but falls short in terms of technical depth and standardization. |
| [18] | 2024 | ✓ | ✓ | ✓ | ✗ | ✓ | This review offers a broad analysis of DL-based osteoporosis classification across modalities but provides minimal coverage of technical aspects and task variations. |
| [19] | 2024 | ✗ | ✓ | ✓ | ✓ | ✓ | This modality-rich survey emphasizes preprocessing and segmentation techniques but lacks comprehensive synthesis and clear taxonomy. |
| [20] | 2025 | ✓ | ✓ | ✗ | ✗ | ✗ | This meta-analysis of DL models using panoramic X-rays offers strong diagnostic synthesis but remains confined to a single modality. |
| Our Survey | 2025 | ✓ | ✓ | ✓ | ✓ | ✓ | Our survey provides a structured and modality-aware synthesis of AI-based methods for osteoporosis detection, bridging clinical tasks with recent DL advances to guide future imaging research. |

## 3. Survey Methodology

This survey provides a systematic and comprehensive review of state-of-the-art AI-based approaches for the detection, prediction, and diagnosis of osteoporosis using medical imaging. To ensure rigor and relevance, we included original research studies published since 2015 that applied AI techniques to one or more medical imaging modalities, namely, X-ray, CT, DXA, and MRI. The scope of inclusion was limited to studies employing AI methods, such as ML, DL, or explainable AI, focusing on diagnostic classification, fracture risk prediction, or related clinical tasks.

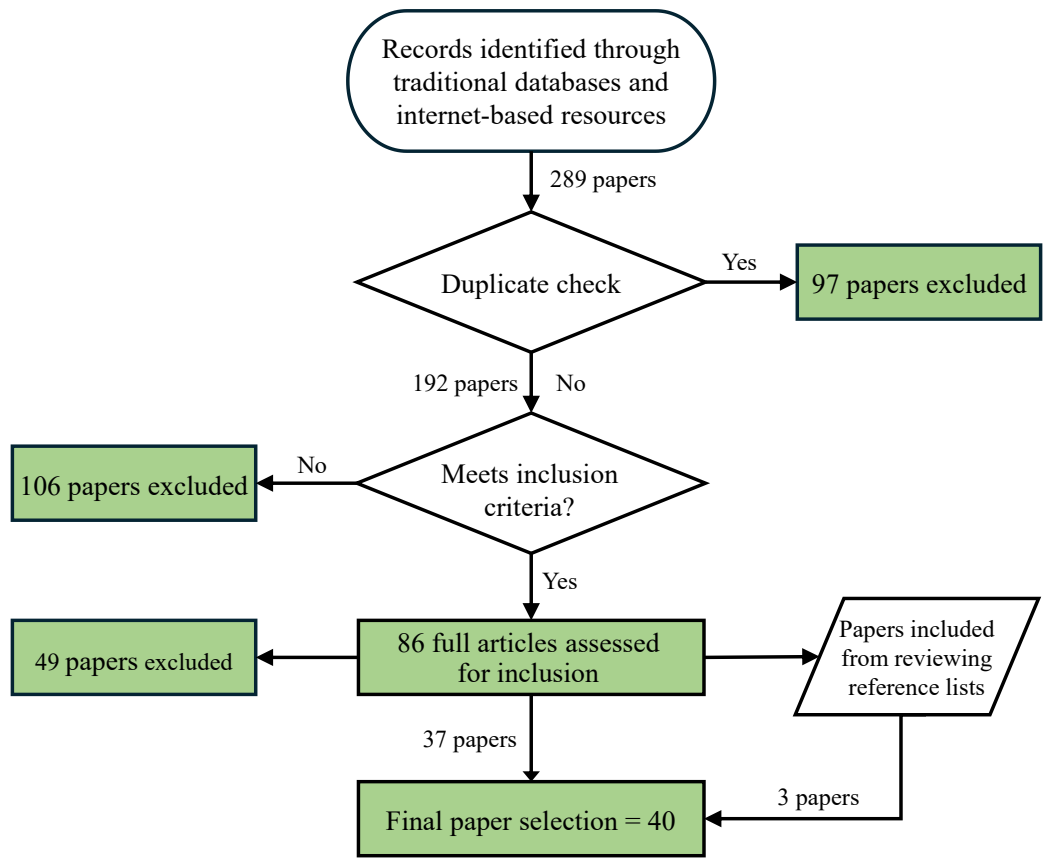

**Figure 1.** Flowchart summarizing the systematic search and selection process for AI-based studies in osteoporosis imaging.

We conducted a systematic and structured literature search to identify relevant AI-based studies on osteoporosis detection, prediction, and diagnosis using medical imaging. The inclusion criteria were restricted to original research articles published from 2015 to June 2025 involving ML, DL, or explainable AI methodologies applied to the X-ray, CT, DXA, or MRI imaging modalities.

To ensure comprehensive coverage of both clinical and technical domains, we searched four major academic databases—PubMed, IEEE Xplore, Web of Science, and Google Scholar—using keyword combinations such as "osteoporosis," "fracture risk," "deep learning," "machine learning," and modality-specific terms. Boolean logic was applied to maximize the search precision. The search was limited to publications between 2015 and 2025. While Google

Scholar was used to complement the indexed databases, its limitations in terms of reproducibility were considered.

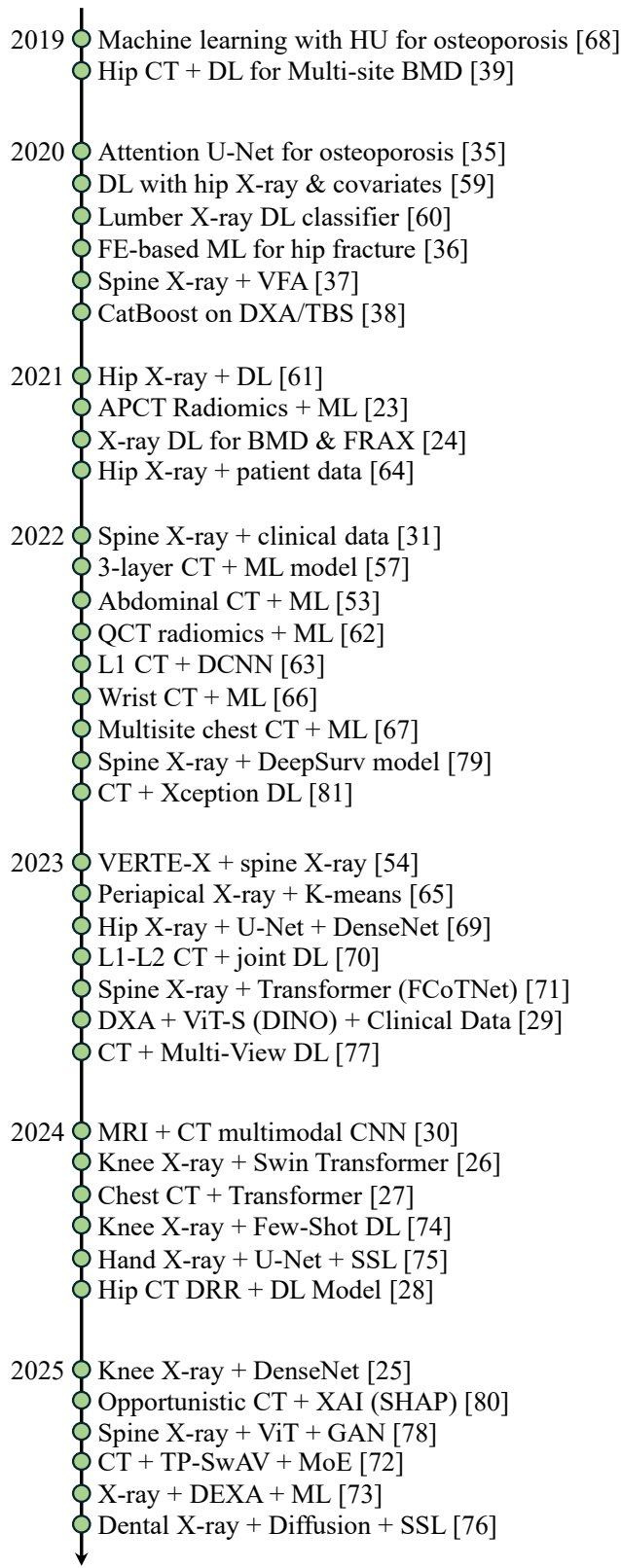

**Figure 2** The timeline chart indicates the year in which each study was made available.

The initial phase of study selection began with screening titles and abstracts to remove duplicates and excluding papers that clearly did not meet the inclusion criteria. Studies were shortlisted if they involved the application of AI to medical imaging for osteoporosis-related tasks, such as detection, diagnosis, classification, segmentation, or prediction of fracture risk. A full-text review was then conducted on the remaining articles to confirm their methodological eligibility. Studies were excluded if they did not use imaging data, lacked an AI component, relied solely on traditional statistical models, or failed to report critical methodological details, such as dataset characteristics, model structure, or evaluation metrics. To ensure that no important studies were overlooked due to search algorithm limitations or indexing delays, we also manually reviewed the reference lists of the included papers. This backward snowballing approach helped to identify additional relevant studies that may not have appeared in the initial keyword-based database search, thereby strengthening the comprehensiveness and representativeness of the final corpus. The full screening process is summarized in **Figure 1**.

We extracted key metadata from each selected study, such as publication year, imaging modality, AI methodology (categorized into classical ML, DL, or explainable AI), dataset type and size, task type (e.g., classification, regression, or segmentation), and metrics used to evaluate performance. In addition to quantitative extraction, we performed a thematic synthesis of the selected literature to identify common research directions, technical strengths, limitations, and clinical applicability. We also recorded whether external validation was conducted, whether public or private datasets were used, and whether model interpretability was considered, all of which are critical factors in assessing clinical readiness and scientific robustness. **Figure 2** provides a timeline of all the studies, showing that early work focused on single-modality approaches with classical ML and basic DL for BMD estimation and osteoporosis classification. From 2021 onwards, studies adopted diverse modalities, incorporated patient data, and applied advanced architectures such as radiomics, CNNs, and survival models. Recent years have shown a shift towards Transformers, multimodal fusion, self-supervised learning, and explainable AI, reflecting a trend towards more robust, interpretable, and clinically applicable solutions.

By adhering to this structured methodology, our survey provides a reliable and nuanced understanding of how AI is being applied to osteoporosis detection through medical imaging. It not only outlines current capabilities and gaps but also lays groundwork for identifying future opportunities where AI can enhance early diagnosis, risk stratification, and personalized bone health management.

***Taxonomy and organizing frameworks***: To provide a structured overview of the literature, we organized the included studies based on a unified taxonomy comprising three core dimensions: imaging modality, clinical task, and AI methodology. This framework allows for a systematic classification of diverse approaches and facilitates meaningful comparisons across studies.

Imaging modalities include DXA, X-ray, CT, and MRI. Clinical tasks are grouped into osteoporosis classification, fracture detection, and risk prediction. AI methodologies are categorized into classical ML, CNN-based models, Transformer architectures, self-supervised learning, and explainable AI. An overview of this taxonomy is illustrated in **Figure 3**, which maps the interconnections between imaging modalities, clinical applications, and AI techniques employed across studies. This multidimensional framework forms the basis for the subsequent sections, where we analyze and synthesize the literature considering these organizing principles, highlighting the current trends, strengths, and research gaps.

The remainder of this survey is organized around this taxonomy. Section 4 presents an overview of the datasets used across the reviewed studies, highlighting their key characteristics and limitations. Section 5 presents a comprehensive analysis of AI-based osteoporosis detection using modality-, task-, and methodology-based structures. Section 6 discusses emerging trends and future directions in the field. Finally, Section 7 concludes the paper with a concise summary of key insights and recommendations for future work.

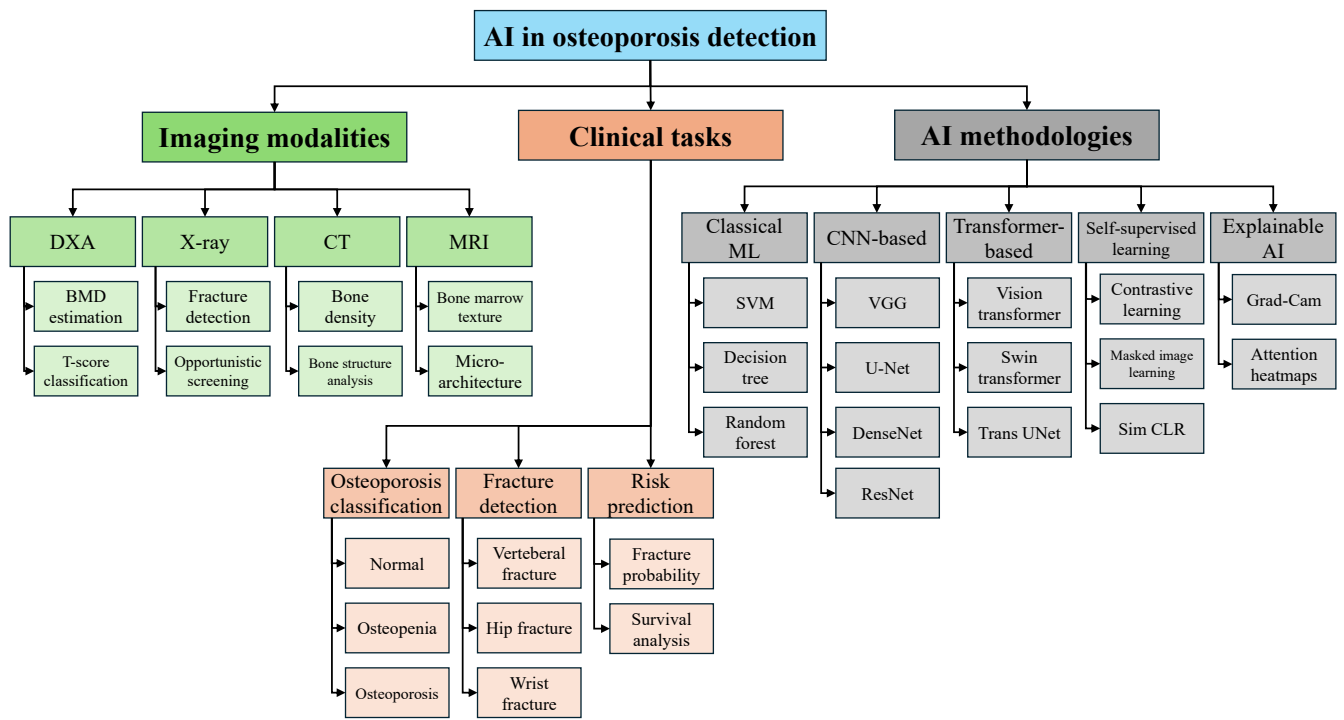

**Figure 3** Overview of AI-based osteoporosis detection across imaging modalities, methodologies, and clinical tasks.

## 4. Datasets

Datasets underpinning recent AI studies on osteoporosis and fracture risk are heterogeneous in terms of scale, access, provenance, and population coverage. To contextualize the reported performance, we profiled the evidence base along practical axes that influence generalizability: public versus private access, single- versus multi-center origin, cohort size, and validation practice. The aggregate picture is summarized in **Figure 1Figure 4**, which visualizes counts and proportions across these dimensions.

Our synthesis shows a pronounced reliance on privately held, single-center cohorts assembled retrospectively. While such datasets enable rapid iteration, they narrow the underlying population and acquisition conditions, increasing the risk that models absorb site-specific biases. External validation remains comparatively uncommon relative to internal resampling; when undertaken, performance frequently attenuates, consistent with a distribution shift between development and evaluation settings. This asymmetry, which is visible in **Figure**

**4**, complicates cross-study benchmarking and limits the strength of claims regarding clinical portability.

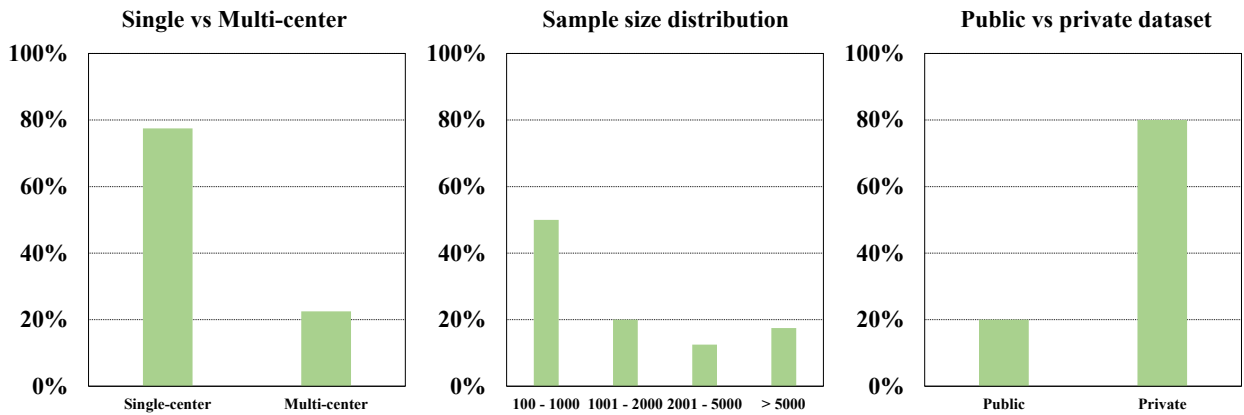

**Figure 4** Dataset landscape across the included studies

Cohort sizes span orders of magnitude, with many studies positioned at the small-to-moderate end of the spectrum. Sparse samples, uneven class prevalence, and limited demographic breadth jointly inflate the variance in summary metrics and weaken statistical precision. Demographic reporting, where available, often reveals narrow inclusion windows, further constraining transportability to distinct care environments and patient mixes.

**Table 2** Summary of datasets

| Dataset Ref | Modality | Sample Size | Center Type | Access |
|---|---|---|---|---|
| [21], [22] | DXA, X-ray | Subjects = 170, images = 795 | Single center | Public |
| [23] | CT | Subjects = 500, images= 500 | Single center | Public |
| [24] | X-ray, DXA | Subjects = 41898, images = 41898 | Multi-center | Partially public |
| [25] | X-ray | Subjects = N/A, images = 372 | N/A | Public |
| [26] | X-ray | Subjects = N/A, images = 744 | N/A | Public |
| [27] | X-ray | Subjects = N/A, images = 2558 | N/A | Public |
| [28] | CT | Subjects = 1480, images = 1480 | Single center | Private |
| [29] | DXA | Subjects = 526, images = 1488 | Single center | Private |
| [30] | CT, MRI | Subjects = 220, images = 2157 | Single center | Private |
| [31] | DXA, X-ray | Subjects = 6908, images = 13816 | Multi-center | Private |

Access constraints remain the principal bottleneck to reuse. Ethical oversight, data-use agreements, and de-identification pipelines are essential for patient protection but slow or preclude open release. Where publicly accessible resources are provided, secondary analyses, ablations, and head-to-head comparisons accelerate, and methodological progress compounds. To strengthen the evidence base, future efforts should prioritize multi-center curation with harmonized inclusion criteria, publish dataset "cards" with clear cohort descriptors and versioning, and pair internal assessments with out-of-distribution testing as standard practice. A concise overview of the key datasets cited in this survey, covering modality, sample size, center type, and access status, is provided in **Table 2**.

**5. AI-based Osteoporosis Detection**

AI has shown immense potential for extracting clinically relevant insights from medical images for osteoporosis-related applications. By leveraging large-scale imaging data, AI models have improved the identification of subtle bone changes, enabled individualized risk stratification, and supported automated decision-making in clinical settings. This section describes how imaging, tasks, and methods intersect during the development of AI-based osteoporosis solutions.

*5.1. Modality-based Analysis*

Accurate and early diagnosis of osteoporosis and related fracture risks relies heavily on medical imaging. Various imaging modalities provide distinct structural and compositional insights into bone tissue, making them essential components of clinical workflows and AI-driven analysis. This survey focuses on four primary imaging modalities commonly used in osteoporosis detection, fracture risk prediction, and bone quality assessment: DXA, X-ray radiography, CT, and MRI. As illustrated in **Figure 5**, the X-ray and CT modalities dominate the published literature on osteoporosis imaging, whereas DXA and MRI, although clinically significant, appear less frequently in AI-based research. Each modality differs in terms of imaging principles, clinical applications, resolution, radiation exposure, and suitability for AI model development. A comparative summary of their strengths and limitations is presented in **Table 3**, providing a clear overview of their diagnostic capabilities and technical limitations. In the following subsections, we briefly introduce each modality and describe its role in osteoporosis diagnosis and fracture risk assessment.

**Table 3.** Summary of imaging modalities for osteoporosis assessment

| Source | Strengths | Limitations |
|---|---|---|
| DXA | <ul><li>Gold standard for BMD measurement</li><li>Low radiation exposure</li><li>Quick and widely available</li></ul> | <ul><li>Limited to two-dimensional (2D) projections</li><li>Cannot assess detailed bone microarchitecture</li><li>Limited to specific anatomical sites</li></ul> |
| X-ray | <ul><li>Inexpensive and widely accessible</li><li>Useful for vertebral fracture detection</li><li>Rapid and simple procedure</li></ul> | <ul><li>Low sensitivity to early-stage osteoporosis</li><li>Cannot measure BMD</li><li>Overlapping anatomy reduces clarity</li></ul> |
| CT | <ul><li>High-resolution 3D imaging</li><li>Quantifies both trabecular and cortical bone</li><li>Enables opportunistic screening</li></ul> | <ul><li>Higher radiation dose</li><li>Not commonly used for routine osteoporosis screening</li><li>Higher cost</li></ul> |
| MRI | • Superior soft tissue and marrow assessment<br>• No radiation exposure<br>• Provides microarchitectural insights | • Expensive and less available<br>• Longer acquisition times<br>• Limited clinical standardization for osteoporosis |

### 5.1.1. DXA

DXA is the most widely used imaging modality for osteoporosis diagnosis and fracture risk assessment. It operates by emitting two X-ray beams of different energy levels through the bone to estimate BMD, which is then used to classify bone health according to WHO criteria [32]. Measurements are typically performed on the lumbar spine, hip, and forearm, which are clinically important sites for osteoporotic fractures.

In clinical practice, BMD values are reported as T-scores and Z-scores, providing a quantitative basis for diagnosis and treatment decisions [4][33]. While axial skeleton sites remain the standard regions of interest for DXA measurements, several studies have explored the feasibility of assessing peripheral BMD [34].

Beyond clinical diagnosis, DXA-based measurements have increasingly been incorporated into AI-based studies for osteoporosis classification and fracture risk prediction. Recent studies have employed ML models trained on clinical and biomechanical features derived from DXA images, finite element analyses, and side-fall simulations to achieve better prediction performance than BMD alone [35][36][37][38][39].

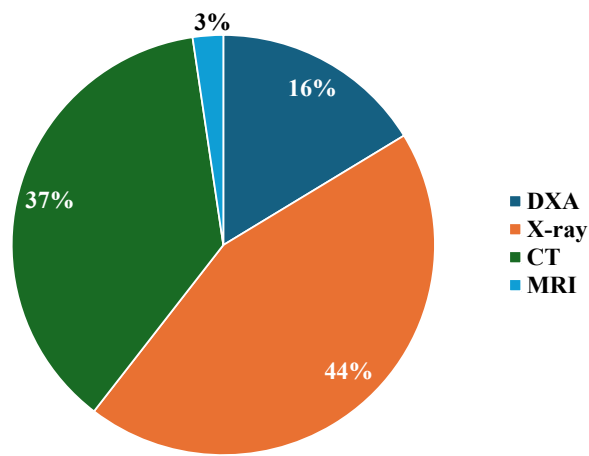

**Figure 5** Distribution of included papers by imaging modality

### 5.1.2. X-ray

X-ray radiography has long been a cornerstone of musculoskeletal assessment and remains one of the most widely used imaging modalities. Although it cannot directly quantify BMD, it plays a crucial role in the detection of osteoporotic vertebral fractures, which are often the first clinical manifestation of osteoporosis [40]. Anteroposterior and lateral spinal radiographs enable the identification of key structural changes, such as vertebral height loss and cortical thinning, using semi-quantitative grading systems like the Genant scale [41].

In addition to conventional assessments, Digital X-ray (DXR) has shown promise in deriving quantitative bone metrics from standard hand or wrist radiographs. DXR estimates the cortical bone thickness and converts it to a BMD proxy, with a precision comparable to that of DXA [42]. This method has demonstrated potential for fracture risk prediction and may serve as a cost-effective adjunct or alternative to DXA in certain settings.

More recently, AI-based techniques have significantly enhanced the analytical capabilities of X-ray imaging for the diagnosis of osteoporosis. For example, CNNs have been trained on large-scale spine radiograph datasets to automatically detect vertebral compression fractures (VCFs) with high diagnostic accuracy [43]. ML models have also been applied to X-ray-derived bone texture and clinical data to improve osteoporosis classification and fracture risk stratification beyond traditional methods [18].

Although X-ray lacks the density metrics provided by modalities such as DXA or CT, its ubiquity, speed, and low cost, combined with AI applications, make it a powerful tool for osteoporosis screening, particularly fracture detection and automated radiographic analysis.

*5.1.3. CT*

CT offers high-resolution, three-dimensional bone imaging, enabling volumetric BMD (vBMD) measurement and detailed evaluation of the trabecular and cortical compartments. Quantitative CT (QCT) assessments of the spine and hip serve as more accurate indicators of fracture risk than DXA [44]. Moreover, opportunistic screening, analyzing routine chest, abdominal, or pelvic CT scans, has proven to be effective in identifying individuals with low vBMD without additional radiation exposure [45][46].

Beyond BMD quantification, CT facilitates advanced analysis through finite element analysis (FEA), which simulates mechanical stress scenarios to estimate bone strength and fracture susceptibility. These FEA-derived parameters have been shown to improve fracture risk discrimination compared with BMD alone [47]. AI has further enhanced CT's diagnostic potential. DL models can automatically segment the vertebrae, perform opportunistic vBMD measurements, and detect vertebral fractures with high accuracy [48][49].

Despite its strengths in structural assessment, CT's higher radiation dose, cost, and limited availability for routine osteoporosis screening limit its use as a first-line diagnostic tool. Nonetheless, CT, including opportunistic analysis, remains invaluable when DXA results are inconclusive or when a detailed fracture evaluation is required.

*5.1.4. MRI*

MRI delivers a high-resolution, radiation-free evaluation of the bone and surrounding soft tissues, enabling an in-depth assessment of trabecular microarchitecture and bone marrow composition, both of which are key factors in osteoporosis beyond mineral density alone [50]. Techniques such as proton density-weighted imaging, T1-weighted sequences, and high-resolution peripheral MRI (HR-pQCT) allow in vivo analysis of trabecular structure and marrow adiposity, which are both associated with bone fragility [51].

Recent MRI approaches have focused on quantifying bone marrow fat and its clinical significance. Marrow adipose tissue (MAT) can be imaged using spectroscopy and multi-echo techniques, which have shown a negative correlation with BMD and positive association with fracture risk [52]. These findings suggest that MRI-derived marrow metrics may serve as novel biomarkers for the management of osteoporosis.

In AI applications, ML and DL models have leveraged MRI-derived texture and microstructural features, bone marrow fat content, and clinical data to classify osteoporosis. Recent DL models have demonstrated high accuracy in classifying osteoporosis using lumbar spine MRI either alone or in combination with CT [30].

Despite its advanced imaging capabilities, MRI's limited availability, higher expense, and longer acquisition times compared with DXA and X-ray primarily limit its use to research settings and select clinical cases where detailed structural or marrow evaluation is warranted.

*5.2. Task-based Analysis*

AI methods have been applied in osteoporosis research to address distinct clinical tasks that support early diagnosis, fracture prevention, and patient management. These tasks include osteoporosis classification, fracture detection, and prediction of fracture risk. Each of these tasks plays a vital role in clinical decision-making and patient care, leveraging various imaging modalities and AI techniques to improve diagnostic accuracy and predictive performance. **Figure 6** indicates that most studies have focused on osteoporosis classification, while relatively fewer have addressed fracture detection and risk prediction.

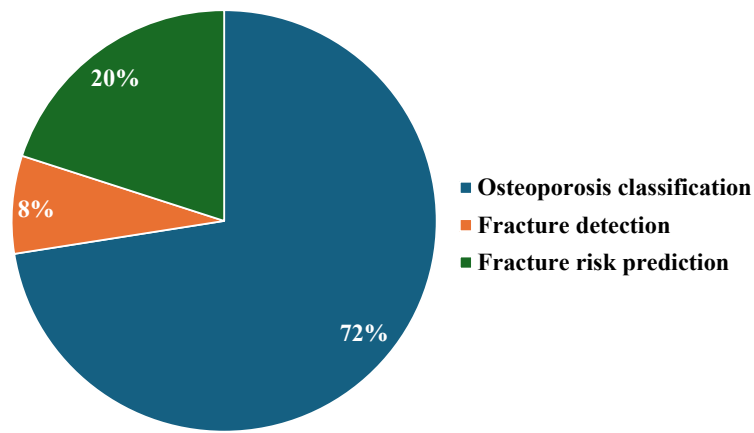

**Figure 6** Distribution of included studies by clinical task

*5.2.1. Osteoporosis Classification*

Osteoporosis classification refers to the categorization of bone health status, typically distinguishing between normal BMD, osteopenia, and osteoporosis. Traditionally, this classification has relied on BMD measurements obtained via DXA and interpreted according to WHO criteria. However, AI-driven approaches have expanded beyond BMD by integrating texture analysis, radiomics, and image-based DL features extracted via X-ray, CT, and MRI scans.

ML algorithms, such as Support Vector Machines (SVM) and Random Forest (RF), and DL models, such as CNNs, have been employed to classify osteoporosis using trabecular patterns, cortical thickness, and other bone quality indicators. These models have shown potential for detecting osteoporotic changes even when BMD is within normal or near-normal ranges, enhancing early diagnosis and reducing reliance on manual interpretation [53][25].

*5.2.2. Fracture Detection*

Fracture detection focuses on identifying existing fractures, particularly those associated with osteoporosis, such as VCFs, hip fractures, and wrist fractures. Early detection of these fractures is crucial, as they often go unnoticed in routine clinical assessments, especially vertebral fractures, which can be asymptomatic.

AI models have been widely used to automate fracture detection in radiographs, CT scans, and MRI. DL methods, particularly CNN-based architectures, have demonstrated superior performance in detecting subtle fracture patterns compared with traditional rule-based or manual assessment techniques. Recent studies have applied DL models to lateral spine and pelvic radiographs to detect vertebral and hip fractures, showing improved diagnostic performance over clinical models [54]. Explainable AI techniques, such as heat maps and Gradient-weighted Class Activation Mapping (Grad-CAM) [55] visualizations have been integrated to highlight fracture sites and provide interpretable outputs that support radiologists in their clinical workflows.

*5.2.3. Fracture Risk Prediction*

Fracture risk prediction involves estimating the probability of future fractures in patients with or without diagnosed osteoporosis. Traditional tools, such as FRAX, consider clinical risk factors and BMD [56]; however, they do not fully account for the structural and compositional bone features visible in medical images. AI approaches have enhanced fracture risk prediction by incorporating image-based biomarkers, such as texture features, cortical porosity, and bone microarchitecture, from DXA, X-ray, and CT scans.

ML algorithms (e.g., logistic regression (LR) and ensemble methods) and DL models (e.g., survival analysis networks) have been used to develop predictive models that forecast fracture occurrence over specified follow-up periods. Longitudinal studies integrating baseline imaging and clinical data have further refined risk prediction by enabling personalized assessments. These AI-based risk models have the potential to complement or even surpass conventional clinical tools by providing more nuanced and individualized risk estimations.

*5.3. Method-based Analysis*

AI methodologies have played a transformative role in analyzing complex medical imaging data for osteoporosis-related clinical tasks. Broadly, these methodologies can be categorized as classical ML, DL, and explainable AI approaches. Each category makes a unique contribution to addressing the different challenges in osteoporosis detection, fracture identification, and risk prediction. The distribution of studies based on methodological analysis is illustrated in **Figure 7**.

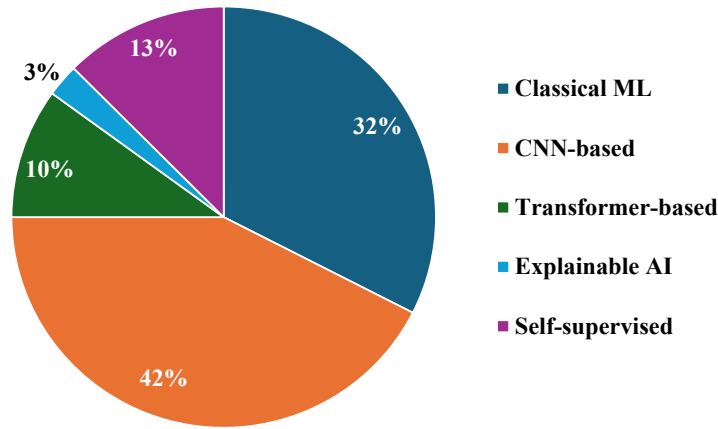

**Figure 7** Distribution of included studies by methodology used

*5.3.1. Classical ML*

Classical ML algorithms have laid the foundation for AI-based osteoporosis assessment. These models rely on handcrafted features extracted from medical images, such as BMD, texture descriptors, and geometric features. Commonly used algorithms include LR, SVM, decision trees, RF, and ensemble learning techniques.

Feature engineering typically involves extracting radiomic features from DXA, X-ray, or CT scans, followed by feature selection and model training. Classical ML methods have been effective for binary osteoporosis classification (normal vs. osteoporosis) and predict fracture risk using structured clinical data combined with imaging biomarkers. However, their reliance on manual feature extraction and limited capacity to capture hierarchical image representations often restrict their performance compared to that of DL models.

Villamor et al. [36] developed a hip fracture prediction model that combines DXA-based 2D finite element (FE) analysis with ML to improve accuracy beyond BMD. DXA scans from 137 postmenopausal women were processed through a multi-step FE workflow (**Figure 8**), beginning with semi-automated segmentation and geometry/density preprocessing, followed by meshing, BMD-based material mapping, and loading/boundary condition (BC) assignment informed by clinical data. The FEBio solver then computed displacements, stresses, and strains, which were post-processed into biomechanical attributes, such as the load-to-strength ratio and site-specific stress-strain metrics. Alongside geometric and clinical variables, 19 low-correlation predictors were trained using multiple classifiers; an SVM with an RBF kernel achieved the highest test accuracy (78.35%), outperforming both BMD-only and clinical-only models. Technically, integrating physics-based FE descriptors with nonlinear SVM leverages the structural and loading information absent in BMD, enabling better fracture risk discrimination in small datasets. However, the inability of the 2D FE model to capture full 3D bone anisotropy, combined with its limited cohort size, may constrain generalization.

Liu et al. [57] proposed a hierarchical opportunistic osteoporosis screening model that integrates clinical data with lumbar spine CT-derived texture features, enabling osteoporosis detection without the need for dedicated DXA scans. From a retrospective cohort of 246 individuals with both modalities available, 35 clinical variables (demographics and laboratory results) and CT-based texture descriptors from manually segmented L1–L4 vertebrae were extracted and filtered through statistical analysis to remove redundant features. The workflow comprised data acquisition, feature extraction, feature selection, and classification using six ML algorithms across three independent layers: demographics only, full clinical data, and combined clinical plus CT texture features. LR achieved the highest area under the ROC curve (AUC), with improved performance as richer feature sets were incorporated. Texture analysis, particularly the Gray-Level Co-occurrence Matrix (GLCM) and histogram features, effectively captured trabecular microarchitectural changes, such as increased inter-trabecular spacing and cortical thinning, which are not reflected in BMD but are strongly linked to fracture risk. The hierarchical design mirrors real-world data availability, allowing the opportunistic use of existing CT scans, while the choice of LR leverages its robustness in small datasets and clinical interpretability. Limitations include reliance on 2D sagittal slices, which cannot fully capture 3D bone architecture, and data from a single center, which may constrain generalization. However, the study demonstrated the potential of combining opportunistic imaging biomarkers with routine clinical data for early osteoporosis screening.

**Table 4.** Summary of included AI-based studies on osteoporosis classification, fracture detection, and fracture risk prediction (AUC: Area under the ROC curve, Acc: Accuracy, Sens: Sensitivity, Spec: Specificity, F1: F1-score).

| Ref. | Method | Dataset | Modality | Performances |
| --- | --- | --- | --- | --- |
| **Osteoporosis classification** | | | | |
| [57] | Classical ML (LR, SVM, RF, ANN, XGBoost, Stacking) | 2210 patients, private | CT | AUC=0.962, Acc=96%, Sens=91%, Spec=92% |
| [58] | Classical ML (Random Forest) | 500 patients, public | CT | Acc=92.7%, AUC=0.96, Sens=80%, Spec=95.8% |
| [53] | Classical ML (LR, SVM, RF, GNB, GBM, XGB) | 172 patients, private | CT | AUROC=0.86, Acc=81%, Sens=70%, Spec=92% |
| [35] | CNN-based (Modified U-Net + Attention) | 126 DXA, 78 X-ray images, private | X-ray, DXA | Acc=88%, Sens=99.7%, Spec=98% |
| [59] | CNN-based (EfficientNet-b3 + clinical covariates) | 1131 patients, private | X-ray | AUC=0.937, Acc=88.5%, Sens=88.7%, Spec=88.2% |
| [60] | CNN-based (3-class CNN, AP + LAT) | 1255 patients, private | X-ray | Osteoporosis: AUC=0.767, Sens=73.7%.: Osteopenia: AUC=0.787, Sens=81.8% |
| [61] | CNN-based (VGG16 + NLNN CNN) | 1001 patients, private | X-ray | AUC=0.867, Acc=81.2%, Sens=91.1%, Spec=68.9% |

| Ref | Method | Dataset | Modality | Metrics |
|---|---|---|---|---|
| [31] | CNN-based (DenseNet + clinical covariates) | 6908 patients, private | X-ray | AUC=0.909, Sens=82%, Spec=85% |
| [62] | Classical ML | 635 patients, private | CT | AUC=0.96, Acc=90%, Sens=98%, Spec=75% |
| [63] | CNN-based | 100 patients, private | CT | Acc=95%, Sens=96%, Spec=94% |
| [24] | CNN-based ResNet-34 and VGG-16 | ~36000 X-rays, partial public | X-ray | AUC=0.97, Acc=91.7%, Sens=80.2%, Spec=94.9% |
| [30] | CNN-based | 2041 Patients, private | MRI, CT | Acc=98.9%, Sens=98.6%, Spec=99.2%, AUC=0.989 |
| [64] | CNN-based ResNet | 1699 patients, private | X-ray | Acc=82.2%, AUC=0.9, F1=83%, Spec=82.5% |
| [65] | Classical ML (MLP, Naïve Bayes, Decision Tree) | 102 patients, private | X-ray | Acc=90.48%, Sens=90%, Spec=90.90% |
| [66] | Classical ML (SVM, Random Forest Classifier) | 196 patients, private | CT | Acc=82%, AUC=0.818, Spec=73.9%, Sens=84.4% |
| [67] | Classical ML (SVM, LASSO, Elastic Net, Ridge LR) | 364 patients, private | CT | Acc=90.4%, Sens=94.7%, Spec=75%, AUC=0.864 |
| [68] | Classical ML (Linear Regression) | 70 patients, private | CT | Acc=92.5%, F1=95.4%, AUC=0.9 |
| [69] | CNN-based (U-Net & U-Net++, pre-trained CNNs (VGG16, ResNet50, DenseNet121)) | 134 patients, private | X-ray | Acc=74%, F1=71%, Sens=68%, Spec=80% |
| [25] | CNN-based (DenseNet169 + Attention Model AM) | 372 X-ray images, public | X-ray | Acc=84.72%, F1=86.08%, Sens=82.93%, Spec=87.1% |
| [70] | CNN-based (ResNet-101 with gated attention) | 1048 patients, private | CT | Acc=95.7%, Sens=97%, Spec=97.6%, F1=97.5%, AUC=0.985 |
| [39] | Classical ML (Random Forest) | 615 patients, private | DXA | Sens=94%, Spec=92%, Acc (Femur)=93.9%, Acc (Spine)=98.3%, Acc (Forearm)=97.9% |
| [71] | Transformer-based (FCoTNet) | 439 patients, 878 X-rays, private | X-ray | Acc: 78.3%; Sens: 69.7%; Spec: 88.9%; F1: 69.9%; AUCs: 0.960/0.823/0.872 (internal); Acc: 55.3% (external) |

| Ref | Method | Dataset | Modality | Results |
|---|---|---|---|---|
| [26] | Transformer-based | 744 X-rays, public | X-ray | Acc: 89.4% |
| [72] | Self-Supervised + CNN + Transformer + Explainable (TP-SwAV, MoE, Multi-task Learning) | 4000+ patients, private | CT | F1 (patient-level): 86.3% (Site A), 82.8% (B), 78.6% (C), 83.6% (D) |
| [27] | CNN-based (DenseNet201), Self-Supervised Learning (SimCLR) + Transfer Learning | 1000 X-rays, private | X-ray | Acc: 94.4%; Sens: 95.5%; Spec: 91.8%; F1: 94.6% |
| [73] | Classical ML (RF, SVM, LR) | 7500 images, private | X-ray, DXA | Acc: 92.5% (RF), Sens: 93.4%, Spec: 91.1%, F1: 92.2%; |
| [74] | CNN-based (VGG, ResNet, Xception) + Few-Shot Learning + AutoML | 1378 X-ray, private | X-ray | Acc: up to 74.5%; Sens: up to 80%; Spec: up to 70.9%; |
| [75] | Self-Supervised Learning (SimCLR, SupCon, VICReg, SwAV) + CNN (ResNet-50) | 1346 X-rays, private | X-ray | AUC 0.85, F1 0.68, Acc 81% (SimCLR); |
| [76] | Self-Supervised + Semi-Supervised (Diffusion model + WideresNet + Sine Threshold) | 12399 X-rays, public release planned | X-ray | Acc: 80.1% |
| [77] | CNN-based (ResNet-18, EfficientNet-b0 | 2883 CTs, private | CT | AUC: 0.964; F1: 81.2%; Sens: 81.3%; Spec: 90.7% |
| **Fracture detection** | | | | |
| [54] | CNN-based (EfficientNet-B4 CNN for VF & OP) | 9276 patients, private | X-rays | VF: AUC=0.93, Sens=76%, Spec=94% -- OP: AUC=0.85, Sens=70%, Spec=83% |
| [78] | Transformer-based | 2820 CTs, private | CT | Acc: 93.68%, Recall: 0.90, F1-score: 0.94 |
| **Fracture risk prediction** | | | | |
| [36] | Classical ML (SVM, LR, RF, NN with FE features) | 137 patients, private | DXA | Acc=78% |
| [37] | Classical ML (LR, SVM, DT, RF) | 137 patients, private | DXA | Acc=87.1%, Sens=83.3%, Spec=92.3% |
| [38] | Classical ML (CatBoost) | 2227 patients, private | DXA | AUC (vertebral fracture) = 0.684, AUC (hip fracture) = 0.656, AUC (total fracture) = 0.688 |
| [79] | CNN-based | 1595 patients, private | X-ray | AUC=0.8015, C-index=0.612 |

| [80] | Transformer-based | 269 patients, private | CT | AUC: 0.913 (train), 0.900 (internal), 0.977/0.964 (external); |
| --- | --- | --- | --- | --- |
| [29] | CNN-based (VGG-16, ResNet-50), Transformer-based (ViT-S) + Self-Supervised Learning (MoCo, DINO) | 526 patients, private | DXA | AUC: 74.3%; F1: 63.6% |
| [28] | CNN-based (DenseNet); 2.5D Ensemble | 1480 patients, private | CT | AUC: 0.74 (2, 3, 5 yrs); C-index: 0.73; |
| [81] | CNN-based (Xception) | 173 patients, private | CT | Acc: 83.9%; AUC: 0.883; PR-AUC: 0.71; External test Acc: 81.7% |

Lim et al. [58] developed an ML model that leverages radiomics features from abdominopelvic CT (APCT) scans to predict femoral osteoporosis, aiming to enable opportunistic screening without additional imaging. In a single-center retrospective cohort of 500 adults who underwent both DXA and APCT within one month, the proximal femur was semi-automatically segmented in 3D, and 41 radiomics features describing intensity, geometry, texture, and wavelet transformations were extracted. Features with high reproducibility were retained, and the top predictors were selected using the random forest feature importance. A random forest classifier trained on these features and optimized with cross-validation achieved AUCs of 0.959 and 0.960 on the training and validation cohorts, respectively, with over 92% accuracy (Acc) and >95% specificity (Spec), significantly outperforming conventional Hounsfield Unit (HU) thresholding. This approach highlights how combining high-dimensional 3D radiomics with ensemble learning can capture subtle trabecular and cortical bone changes, although its single-center scope and binary osteoporosis/non-osteoporosis classification may limit its generalizability.

Ref. [53] developed an ML framework to predict osteoporosis from unenhanced abdominal CT scans by analyzing psoas muscle radiomics as a surrogate for DXA-based bone assessment. In a retrospective cohort of 172 adults over 40 years with paired CT and DXA within three months, the bilateral psoas muscles at the L3 vertebral level were segmented in 3D, and 826 radiomics features covering intensity, shape, texture, and wavelet domains were extracted using PyRadiomics. Dimensionality reduction via the Mann-Whitney U test and Least Absolute Shrinkage and Selection Operator (LASSO) [82] retained 16 discriminative features, which were used to train six classical ML classifiers. Gradient Boosting Machine (GBM) [83] achieved the best validation performance (AUC=0.86, Spec=0.92), outperforming other models. This approach leverages the known muscle–bone interaction, where sarcopenia correlates with lower bone strength, allowing for the inference of osteoporosis risk from routinely acquired abdominal CT scans without the need for additional imaging. Using a high-capacity ensemble method like GBM, the model effectively captured subtle texture and morphological patterns in muscle tissue linked to bone fragility. However, its single-center design, modest sample size, and lack of external validation limit its generalizability.

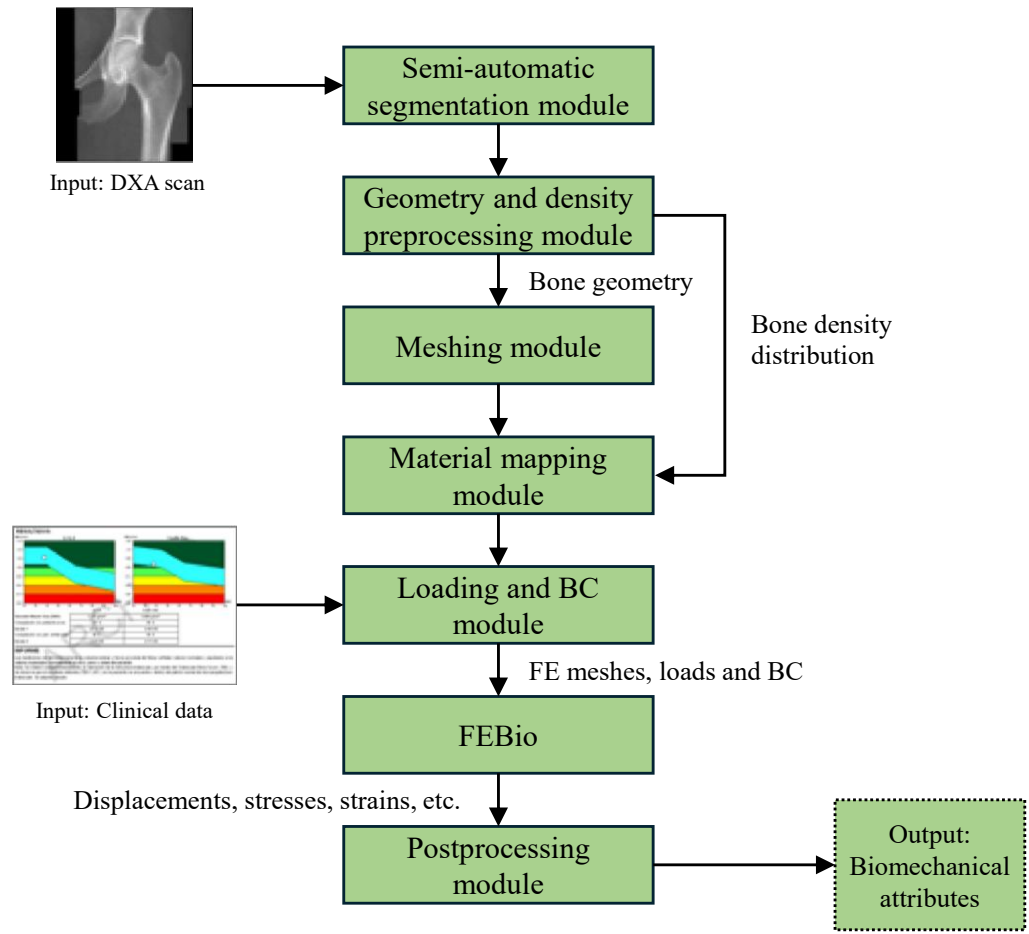

**Figure 8** Workflow of the patient-specific DXA-based FE modeling process for biomechanical attribute extraction, adapted from [36]

In [62], the authors developed and validated a radiomics-based ML model using Quantitative CT (QCT) images of the lumbar spine to distinguish osteoporosis from osteopenia, integrating imaging features with clinical variables to improve stratification. As illustrated in **Figure 9**, in a retrospective cohort of 635 patients from Dazhou Central Hospital, the cancellous bone of L3 was manually segmented, and 851 radiomics features describing intensity, shape, texture, and wavelet domains were extracted. Feature reduction via minimum redundancy–maximum relevance (mRMR) and LASSO retained six discriminative features, which, along with age, alkaline phosphatase, and homocysteine, were used to build a multivariable LR model. The combined model achieved an AUC of 0.96 in both training and test cohorts, outperforming the clinical-only model (AUC = 0.81-0.79). QCT was chosen over DXA because of its ability to separately quantify cortical and trabecular bone density, thus mitigating artifacts from degenerative changes. The incorporation of wavelet-based and first-order radiomics features allowed the model to capture trabecular microarchitectural heterogeneity, while LR offered interpretability and stability on moderately sized datasets. However, the absence of external validation and follow-up outcome data limits the generalizability of the results. This study highlighted the potential of combining high-dimensional radiomics with relevant biochemical markers to refine opportunistic bone health assessments beyond conventional BMD.

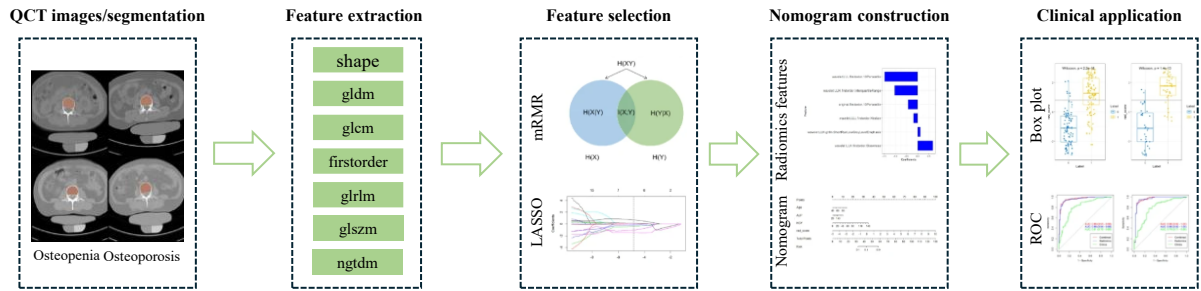

**Figure 9** Flowchart of the radiomics method by [62]

Ref. [37] presented the use of an ML framework for hip fracture risk assessment by integrating clinical, geometric, and biomechanical variables derived from patient-specific DXA-based finite element models simulating sideways falls. Unlike traditional BMD-based screening, which captures bone density but not mechanical response, their approach extracted stress, strain, and fracture risk indices to capture structural integrity under realistic loading. Multiple models were evaluated, with RF achieving the best balance of Sensitivity, Specificity, and Accuracy. The superior performance of RF can be attributed to its ensemble nature, which handles heterogeneous feature types and nonlinear interactions effectively, particularly in small-to-moderate datasets with mixed biomechanical and clinical predictors. In contrast, SVMs showed sensitivity advantages but suffered from poor specificity, likely due to decision boundaries being skewed in imbalanced, high-dimensional spaces, even after synthetic minority oversampling technique (SMOTE) augmentation. LR underperforms because of its linear assumptions, which limit its ability to model complex biomechanical–clinical interactions. This study demonstrated that incorporating finite element-derived mechanical attributes into non-parametric ensemble models can substantially improve fracture risk prediction over density-only approaches, especially when the dataset is enriched with domain-relevant structural predictors.

Ref. [38] developed an ML-based fracture risk model using longitudinal data from 2,227 participants in the Ansung community cohort, integrating DXA-derived BMD, trabecular bone score (TBS), and a wide array of clinical and biochemical variables. Three models CatBoost [84], LR, and SVM, were compared using all features and a reduced top-20 variable set, with SHapley Additive exPlanations (SHAP) applied for interpretability. CatBoost consistently outperformed FRAX and the other ML baselines, achieving superior AUC values in total and hip fracture predictions. Its advantage stems from gradient boosting's ability to model complex nonlinear interactions, handle categorical variables without extensive pre-processing, and resist overfitting in moderately sized datasets. In contrast, SVM underperformed, likely because of its sensitivity to high-dimensional, heterogeneous clinical data, and LR's linear form limited its ability to capture multivariate dependencies. The SHAP analysis revealed both conventional predictors (total hip, lumbar spine, and femur neck BMD) and less traditional factors (arthralgia score, homocysteine, and creatinine) as top contributors, illustrating CatBoost's capacity to identify novel risk markers. However, the absence of external validation and the inclusion of variables that are not routinely collected in clinical practice may limit the real-world generalizability. This study underscored the value of explainable gradient boosting models in fracture risk prediction, particularly when leveraging diverse longitudinal datasets to identify both established and emerging predictors.

In [65], researchers proposed an automated osteoporosis detection pipeline that leverages periapical dental radiographs, with ground-truth labels obtained from DXA scans, targeting opportunistic screening in postmenopausal women. As shown in **Figure 10**, the method focused on ROI-based trabecular bone analysis using pixel-based clustering (K-means and Fuzzy C-means) to segment bone from porous regions, followed by feature extraction via pixel distribution histograms and classification with decision tree, naïve Bayes, and multilayer perceptron (MLP). The best performance (Acc = 90.48%, Spec = 90.90%, Sens = 90.00%) was achieved by combining K-means segmentation with an MLP classifier, a synergy likely due to K-means producing sharper structural separation in homogeneous trabecular patterns, which complements the MLP's ability to model nonlinear pixel-distribution relationships. In contrast, Fuzzy C-means segmentation, while being more tolerant of noise, yielded less distinct boundaries, and naïve Bayes underperformed because of its strong independence assumptions, which do not hold for spatially correlated trabecular features. This approach demonstrated how simple clustering-based segmentation coupled with lightweight neural architectures can deliver high diagnostic accuracy in small datasets without DL, making it suitable for low-resource clinical contexts. However, its single-center dataset and limited sample size restrict its generalizability, underscoring the need for multi-center validation and robustness testing against variable imaging conditions.

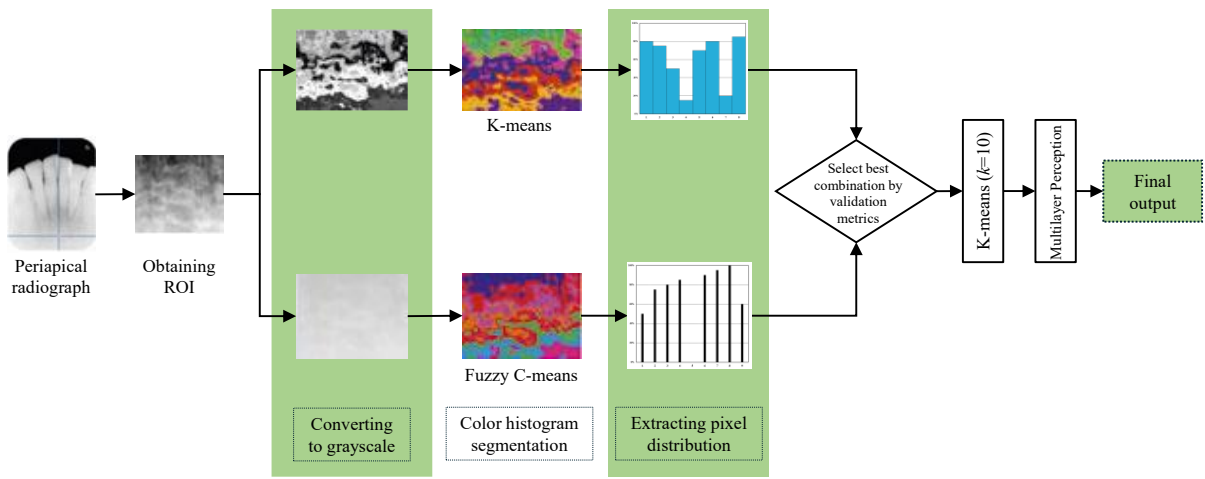

**Figure 10** Schematic of the proposed method in [65]

Ref. [66] developed a DL-based framework to classify normal bone density, osteopenia, and osteoporosis from lumbar spine radiographs, aiming to provide a faster and more cost-effective alternative to DXA. Using 1,201 DXA-labeled lateral lumbar X-rays from a single center in Iran, the authors trained a ResNet50 CNN with transfer learning, applying grayscale preprocessing, data augmentation, and class weighting to address dataset imbalance. While CNNs such as ResNet excel in extracting local spatial features from medical images, their receptive field limitations can restrict the modeling of long-range anatomical dependencies, something that, for instance, Transformer-based architectures could address. However, CNNs retain a strong inductive bias for spatial locality and translation invariance, making them more data-efficient and less prone to overfitting than Transformers when training data are limited, as in this study. The choice of transfer learning leverages pre-trained feature extractors to mitigate the small sample size, improving generalization. Despite the robustness of the method within the dataset, the lack of external validation raises uncertainties regarding its performance on different populations and imaging settings. This highlights a broader methodological challenge: even technically strong models can face reduced generalizability without multi-center, demographically diverse testing, which is critical for clinical translation.

Ref. [67] explored classical ML models to predict osteopenia and osteoporosis using CT attenuation features from multiple osseous sites visible on routine CT scans, leveraging opportunistic imaging data from 364 Mayo Clinic patients with paired DXA within six months. Volumetric 3D segmentation was applied to the ribs, thoracic vertebrae, sternum, and clavicles to extract 36 trabecular attenuation features along with demographic and scan-related variables. Four models—LASSO, ridge regression, elastic net, and Radial Basis Function (RBF) kernel SVM—were trained with an 80:20 split. Classical regression-based methods, such as LASSO and elastic net, offer built-in feature selection and coefficient shrinkage, making them interpretable and well-suited for high-dimensional but moderately sized datasets. However, these models primarily assume linear relationships and may struggle to capture the complex, nonlinear interactions inherent in variations in bone density across multiple skeletal sites. In contrast, the RBF SVM can model such nonlinearities by mapping features into a higher-dimensional space, which yielded superior predictive performance in this study. However, SVMs are more sensitive to feature scaling and hyperparameter tuning, and without abundant, heterogeneous training data, there remains a risk of overfitting, especially when applied to demographically narrow cohorts like this one. While this study demonstrated how routine chest CTs can be repurposed for bone health assessment, the lack of external validation limits generalizability, underscoring the need for multi-center, diverse population testing before clinical adoption.

In [68], an ML framework was developed to predict vertebral T-scores and classify osteoporosis based on HU measurements from conventional lumbar CT scans, using QCT-derived values as the ground-truth. The dataset comprised 198 vertebrae from 70 patients undergoing preoperative spinal surgery, with HU values extracted from three trabecular ROI levels (superior, middle, and inferior) across L1–L3. The authors employed multivariable linear regression to estimate T-scores from age, sex, and mean HU, followed by LR for binary classification. Regression-based models were selected for their interpretability and computational simplicity, enabling clinicians to directly relate input features to the predicted bone density. However, these models assume mostly linear relationships between features and outcomes, which may not accurately capture the complex biomechanical and compositional factors that influence bone strength. LR provides a transparent decision boundary for classification; however, its capacity to model higher-order interactions is limited without engineered feature expansions. The integration of a TensorFlow-based user interface enhanced its usability for clinical adoption, enabling real-time prediction from routine CT data. Nevertheless, the small sample size and manual ROI placement raised concerns about robustness and reproducibility. Automated ROI extraction and larger, multi-center datasets would be needed to improve generalizability. This work demonstrated how accessible and interpretable ML tools can bridge the gap in facilities lacking QCT while highlighting the trade-off between simplicity and modeling capacity in medical imaging applications.

Hussain and Han [39] developed a computer-aided osteoporosis detection (CAOD) system for automated BMD measurement from DXA images across multiple anatomical sites, including the femur, spine, forearm, and femoral implants. The pipeline integrates a non-local means filter for noise suppression, pixel label random forest (PLRF) for tissue segmentation, and contour-based ROI localization to standardize the BMD computation. Non-local means filtering preserves fine bone texture while reducing Gaussian noise, which is critical for accurate segmentation of low-contrast DXA images. The PLRF segmentation approach leverages pixel-level feature learning combined with ensemble decision trees, making it effective for heterogeneous tissue boundaries; however, it remains sensitive to feature quality and class imbalance. Contour-based ROI detection enforces consistent anatomical boundaries, reducing inter-operator variability, which is a known limitation of manual DXA analysis. On a dataset of 615 images, the system achieved 98.4% ROI selection accuracy and reduced variability in BMD estimates compared to manual processing. However, its reliance on classical ML over DL constrains its adaptability to unseen anatomical variations, and certain regions in the study suffered from limited training data, which may have impacted the generalization. Despite these limitations, the CAOD framework demonstrates how algorithmic standardization can enhance the reproducibility and efficiency of DXA-based osteoporosis assessment, particularly in clinical environments that require consistent multi-site analysis.

Ref. [73] developed an ML framework for early osteoporosis detection using X-ray and DXA images sourced from public datasets, including MrOS and Mendeley, containing up to 10,000 labeled cases across normal, osteopenia, and osteoporosis categories. The study performed a binary classification of bone health status through a pipeline of noise reduction, contrast enhancement, segmentation, and feature extraction via wavelet transforms and PCA. Three classifiers, SVM, RF, and LR, were trained using stratified data splits with cross-validation and hyperparameter tuning. Classical ML methods were selected for their interpretability and computational efficiency in resource-constrained settings. RF achieved the highest accuracy (92.5%), likely owing to its ensemble-based ability to capture nonlinear feature interactions and model complex dependencies between trabecular texture and cortical structure, outperforming linear LR and margin-based SVM in this context. However, the reliance on handcrafted features introduces a key limitation: feature design depends heavily on domain expertise and may fail to capture subtle, higher-order imaging patterns that DL could learn automatically. Moreover, the reported computational demands for preprocessing and feature engineering highlight a scalability challenge for large-scale clinical integration. While the framework demonstrates that well-optimized classical models can be effective for opportunistic osteoporosis screening, the authors noted that transitioning towards DL architectures could reduce manual intervention, enhance feature representation, and improve robustness across diverse datasets.

Yet, despite these advances, classical ML methods depend on handcrafted features and heavy pre-processing, making them sensitive to scanner/protocol differences and operator choices. Their limited capacity struggles with 3D bone anisotropy, multi-scale trabecular patterns, and complex imaging-clinical interactions; therefore, models are often reduced to binary tasks on small, single-center cohorts. Feature selection can be unstable with modest sample sizes and is prone to leakage; label heterogeneity (e.g., DXA vendors, HU surrogates, and muscle proxies) and class imbalance further compromise robustness. Physics/opportunistic features add signals, but still miss the full biomechanical context and are tightly coupled with specific workflows. Crucially, external validation, calibration, and prospective decision-curve analyses are rare; therefore, the reported gains may not translate to clinical practice.

*5.3.2. CNN-based Approaches*

CNNs have emerged as a dominant approach in medical image analysis owing to their ability to automatically learn hierarchical features from raw imaging data. CNNs have been widely employed in osteoporosis detection for tasks such as bone segmentation, feature extraction, and classification using X-ray, CT, and DEXA images. Their spatial invariance and capacity to capture both local and global structural patterns make them particularly well-suited for identifying osteoporotic changes in bone microarchitecture. This section reviews CNN-based methodologies developed for osteoporosis diagnosis, highlighting their architecture, training strategies, and performances across different imaging modalities.

In [35], the authors presented a DL-based framework for osteoporosis diagnosis using a modified U-Net architecture with integrated attention units, targeting segmentation of bone regions in both DEXA and X-ray images, as shown in **Figure 11**. The system was trained on two internally collected datasets: DEXSIT (DEXA) and XSITRAY (X-ray), with segmentation serving as the foundation for computing BMD and T-scores. The modified U-Net incorporates residual blocks and both custom channel and spatial attention modules (CBAM) to enhance feature representation. Attention mechanisms enhance segmentation by suppressing irrelevant background features and emphasizing both global structural cues and fine trabecular patterns, addressing the challenge of low-contrast boundaries in bone imaging. Residual connections mitigate vanishing gradients and facilitate deeper feature extraction, while the combination of channel and spatial attention helps to adaptively refine salient features at different receptive field scales. Post-segmentation, a handcrafted mathematical model estimates BMD and T-scores, enabling classification into normal, osteopenia, or osteoporosis categories, with outputs validated against clinical DEXA reports. This pipeline strikes a balance between interpretability and performance, providing clear visual attention maps that foster clinical trust. However, its reliance on a small, internally collected dataset limits generalization, and the handcrafted BMD estimation formula introduces potential bias from population-specific parameters. While this work demonstrated how attention-augmented U-Nets can substantially improve segmentation accuracy in multimodal osteoporosis screening, its dependency on manual calibration and limited data highlight the need for larger, multi-center datasets and end-to-end learning for more robust deployment.

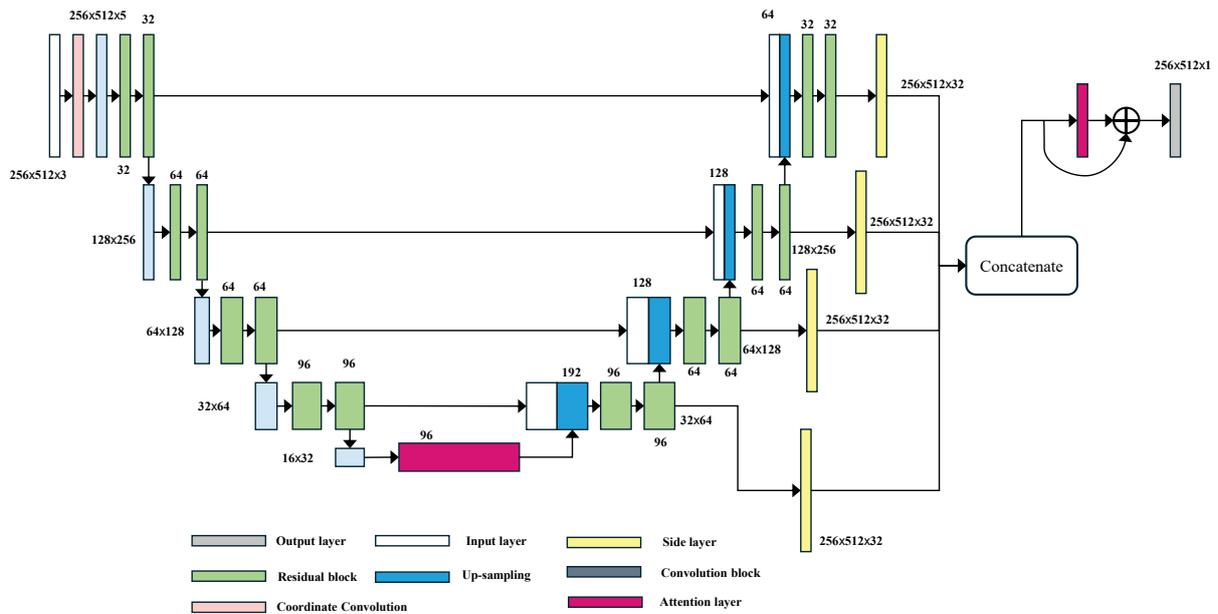

**Figure 11** U-Net architecture proposed in [35]

In [59], a DL framework was developed for osteoporosis classification of hip radiographs, further examining whether structured clinical covariates could enhance image-based predictions. The dataset comprised 1,131 radiographs from patients aged 60 years or older, each paired with DXA-derived T-scores. Five pre-trained CNNs, ResNet18/34, GoogleNet, EfficientNet-b3, and EfficientNet-b4, were fine-tuned on manually cropped hip regions that mimic DXA measurement fields. To integrate clinical context, an ensemble architecture concatenated the CNN-extracted image embeddings with covariates (age, sex, BMI, and fracture history) through a fully connected layer. The best performance was achieved by EfficientNet-b3 with combined inputs (AUC = 0.9374), outperforming the image-only models. This gain can be attributed to the complementary nature of structured clinical variables, which capture systemic risk factors that CNNs cannot directly infer from morphology. EfficientNet's compound scaling and depth–width balance likely contributed to superior feature extraction over shallower or narrower CNNs, especially in capturing fine trabecular and cortical patterns from moderate-sized datasets. Grad-CAM visualizations confirmed focus on anatomically relevant regions near the lesser trochanter, improving interpretability. However, limitations include the narrow set of patient variables, single-center origin, and manual cropping, which may restrict scalability and generalizability. This study highlighted that in opportunistic osteoporosis screening, hybrid models blending radiographic and clinical data can provide a more robust decision framework than either modality alone, particularly when using architectures that maximize feature diversity under limited training data conditions.

A study by Zhang et al. [60] investigated the use of deep CNNs (DCNNs) for opportunistic BMD classification from routine lumbar spine radiographs in postmenopausal women, leveraging multi-center data from three tertiary hospitals in China. The task involved multi-class classification into normal, osteopenia, or osteoporosis categories using DXA-derived T-scores as a reference, with models trained directly on raw anteroposterior and lateral X-ray views as well as their combination. By bypassing handcrafted features, the DCNNs could learn imaging biomarkers that capture both the trabecular texture and macrostructural cues, potentially detecting patterns that are imperceptible to human readers. The multi-view approach allowed the network to integrate complementary anatomical information, anteroposterior views providing a global perspective on vertebral alignment, and lateral views offering better trabecular visibility, although the ROI definitions excluded cortical bone, potentially limiting sensitivity. While the multi-center design enhanced cohort diversity compared with single-institution studies, sensitivity remained relatively low, suggesting that image quality variation and subtle osteopenia features challenged the model's discriminative capacity. Performance metrics, including AUC and likelihood ratios, highlighted the clinical promise of DCNNs as low-cost, radiation-free screening tools where DXA is unavailable. However, the absence of external validation beyond Chinese cohorts raises concerns about population generalizability. This work demonstrated that direct learning from routine radiographs can align with real-world clinical workflows; however, the optimization of sensitivity and cross-population robustness remains essential for broader adoption.

Ref. [61] developed a DL model for binary osteoporosis classification from simple hip radiographs, aiming to provide a practical alternative in settings where DXA access is limited. The dataset consisted of 1,001 cropped proximal femur images from postmenopausal women (T-score $\leq -2.5$ vs. $> -2.5$), split into training, validation, and test sets, with an additional external validation cohort of 117 cases. The core architecture was VGG16, augmented with a nonlocal neural network (NLNN) module to address the inherent limitations of CNNs in capturing only local features. The NLNN introduced a self-attention mechanism that computed pixel–pixel correlations, enabling the network to model global contextual relationships such as long-range cortical continuity and diffuse trabecular loss, which are critical for BMD estimation from plain radiographs. Preprocessing involved Z-score normalization with randomized scaling to handle nonstandard acquisition parameters and extensive augmentation to improve robustness. The model achieved high internal sensitivity (91.1%) but a notable drop in external performance (AUC from 0.867 to 0.700), underscoring domain generalization challenges in medical imaging. Grad-CAM visualizations confirmed a physiologically plausible focus on the proximal femoral cortex and trabecular architecture, supporting its clinical interpretability. However, the binary classification design excluded osteopenia, potentially oversimplifying the continuous spectrum of bone loss, and the restriction to hip images omitted complementary data from the spine. While the study demonstrated that integrating global context modeling into CNNs can enhance feature extraction for opportunistic osteoporosis screening, its real-world adoption will depend on improving the cross-domain robustness and expanding the task to multi-class or continuous BMD prediction for richer clinical utility.

Ref. [54] proposed a DL-based approach for the simultaneous detection of osteoporosis and vertebral fractures using lateral spine X-rays, aiming to improve opportunistic screening and DXA referral rates. The study utilized 26,299 images from 9,276 individuals for training and testing and an external cohort of 398 images for validation. The objective was to develop two DCNN models, VERTE-X pVF (for vertebral fractures) and VERTE-X osteo (for osteoporosis), based on EfficientNet-B4, trained on whole radiographs without region-specific cropping. The evaluation metrics showed high performance (AUROC: 0.93 for VF and 0.85 osteoporosis internally; 0.92 and 0.83 externally), outperforming clinical models based on structured risk factors. Grad-CAM visualizations confirmed attention on vertebral regions, enhancing interpretability. The strengths include large-scale, multi-vendor data, external validation, and dual-outcome prediction. The limitations include restricted generalizability beyond Korean populations and omission of anterior-posterior views. Overall, the VERTE-X system shows promise for integrating AI into routine radiography to enhance the identification of early fractures and osteoporosis risk.

Mao et al. [31] proposed a DenseNet-based CNN for multi-class classification of normal BMD, osteopenia, and osteoporosis from paired anteroposterior (AP) and lateral (LAT) lumbar spine radiographs, using DXA-derived BMD as the reference standard. The dataset comprised 6,908 individuals aged 50 years or older from multiple Chinese centers, enabling broader demographic coverage than most prior single-site studies. The model employed a dual-channel architecture to process AP and LAT images in parallel, with feature maps fused before classification. This design leveraged the complementary strengths of the two views: AP images capturing overall vertebral shape and cortical outlines, and LAT images offering better visualization of trabecular microarchitecture. As illustrated in **Figure 12**, the extensive preprocessing standardized image resolution and contrast across centers, mitigating variability from heterogeneous imaging equipment. The study also compared image-only models with hybrid models that concatenated learned image features with clinical covariates (age, sex, and BMI) prior to the classification layer. The hybrid approach improved performance, reflecting the additive diagnostic value of systemic risk factors that cannot be inferred purely from morphology. DenseNet's dense connectivity promotes feature reuse and gradient flow, making it particularly effective for learning fine-grained bone texture patterns in large, heterogeneous datasets. However, the sensitivity for osteopenia remained lower than that for osteoporosis or normal classes, likely due to subtler imaging differences and overlap in BMD distribution. Additionally, excluding younger individuals limited the generalizability to early detection scenarios. Overall, this study demonstrated how multi-view, multimodal integration can enhance opportunistic osteoporosis screening from routine radiographs, particularly in large-scale, multi-institutional settings.

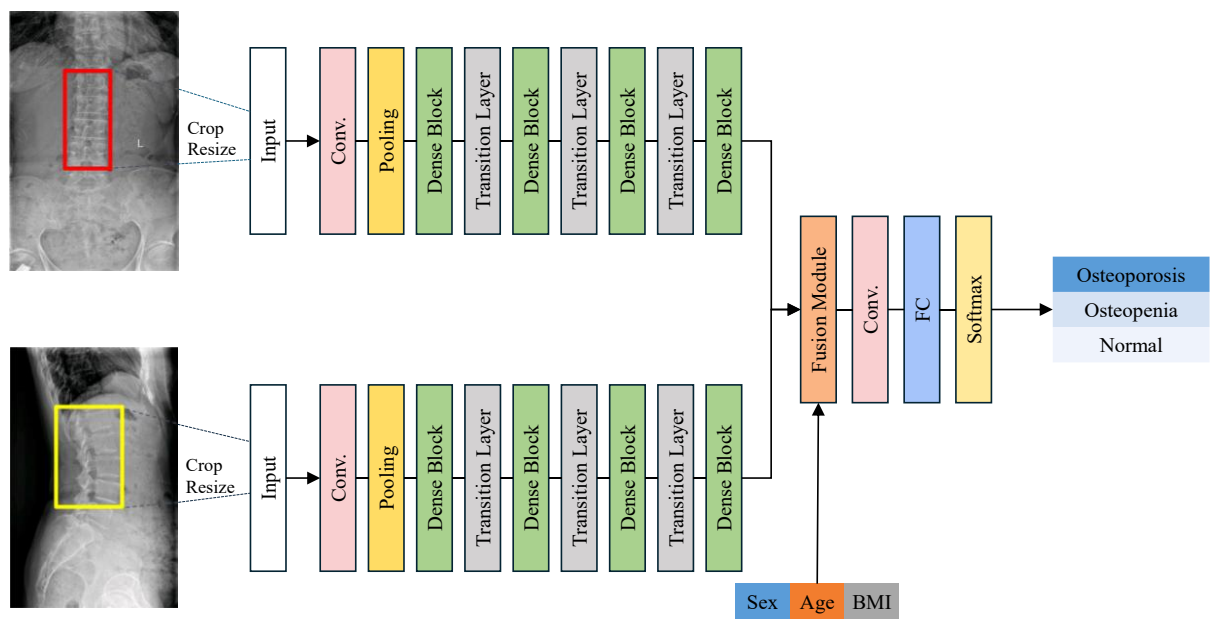

**Figure 12** Overall proposed framework by Mao [31]

Dzierżak and Omiotek [63] explored the use of DCNNs for binary classification of osteoporosis from CT images of L1 vertebral trabecular bone. The dataset consisted of 400 manually cropped 50×50-pixel patches from 100 patients (50 healthy and 50 osteoporotic), focusing on the spongy tissue region to capture microstructural patterns associated with bone fragility. Six pre-trained CNN architectures–VGG16, VGG19, MobileNetV2, Xception, ResNet50, and InceptionResNetV2–were fine-tuned using transfer learning and data augmentation to mitigate the overfitting risks posed by the small dataset. VGG16 achieved the highest accuracy (95%), true positive rate (96%), and true negative rate (94%), which can be attributed to its relatively shallow depth and smaller parameter count compared with deeper networks, allowing better generalization when training data are scarce. In contrast, more complex architectures, such as InceptionResNetV2, may overfit due to their higher capacity and data-hungry nature. Manual ROI extraction ensured anatomical consistency but introduced potential operator bias and limited scalability, and automated localization methods could improve reproducibility in clinical deployment. Although the study was constrained by its small, single-center dataset and artificially cropped inputs, it highlighted that lightweight pre-trained CNNs can deliver high diagnostic performance for CT-based osteoporosis detection under data-limited conditions, provided that careful model selection and augmentation strategies are applied.

In [24], a DL-based tool was developed to predict BMD and assess fracture risk from plain pelvic and lumbar spine radiographs, enabling opportunistic screening without additional imaging. The model was trained on over 36,000 radiographs and DXA pairs, using automated anatomical landmark detection and image quality assessment to standardize inputs prior to BMD regression. Two architectures, VGG16 and ResNet34, were evaluated, with the latter's residual connections improving gradient flow and feature reuse, making it more effective for capturing both fine trabecular details and broader cortical patterns. The predicted BMD values were subsequently integrated into the FRAX tool to estimate 10-year fracture risk, providing a clinically interpretable pathway from image acquisition to risk stratification. High correlations with DXA-derived BMD ($r = 0.92$ for hip, $r = 0.90$ for spine) and strong AUROC values (0.97 for hip, 0.92 for spine) demonstrated the framework's accuracy. The use of fully automated landmark detection reduced operator variability and allowed scalability to population-level screening. However, performance remained dependent on image quality, and the model incorporated a limited clinical patient history, potentially omitting non-imaging risk factors that are critical for individualized fracture prediction. This work demonstrated that, when coupled with robust preprocessing and large-scale training, DL applied to widely available radiographs can deliver clinically actionable BMD and fracture risk assessments, supporting integration into routine healthcare workflows.

In [30], a CNN model was proposed using both unimodal and multimodal features to predict osteoporosis based on lumbar MRI and CT scans. The dataset included MRI and CT images from 246 patients with low BMD and 108 healthy controls, with BMD measured via DEXA as a reference. The goal was to distinguish between osteoporotic and normal bone structures using MRI, CT, or both modalities together. The authors designed a custom CNN (see **Figure 13**) with parallel dual blocks (3×3 and 5×5 convolutions) for richer feature extraction, and they compared this architecture with traditional CNNs and six pre-trained models. An extensive evaluation using five experimental stages and metrics, including accuracy, sensitivity, specificity, balanced accuracy, and ROC-AUC, demonstrated that the proposed multimodal model outperformed all others, achieving a balanced accuracy of up to 98.90%. The strengths of this study included its architectural innovation, rigorous testing (including a hold-out test and patient-based analysis), and potential to reduce radiation exposure by effectively utilizing MRI. However, its limitations included the lack of external validation and exclusion of patients with normal BMD from the control group. Overall, the work highlighted the value of multimodal CNNs for the accurate prediction of osteoporosis across various imaging modalities.

Yamamoto et al. [64] evaluated whether combining clinical variables with hip X-ray images could enhance DL-based osteoporosis classification. The dataset consisted of 1,699 hip radiographs from patients aged 60 and above, labeled using DXA-derived T-scores. Osteoporosis was defined as a T-score ≤ −2.5. The authors used five ResNet architectures (18–152 layers) with transfer learning and trained both image-only CNN models and ensemble models that incorporated patient data (age, sex, and BMI). The ensemble model combined features extracted from the CNN with clinical variables through a fully connected layer. Performance was assessed using accuracy, AUC, precision, recall, specificity, and F1 score across a 4-fold cross-validation repeated 30 times. The results showed that adding patient variables to image-only models led to statistically significant improvements in AUC, especially in the ResNet34 model (AUC increase of 0.004, $p = 0.0004$, effect size = 0.871). A key strength of this study was the statistical comparison between models and the real-world applicability of the selected variables. However, its limitations included manual cropping, limited clinical features, and a lack of external validation. The study highlighted that integrating simple clinical information into CNN models improves osteoporosis classification and supports real-world deployment.

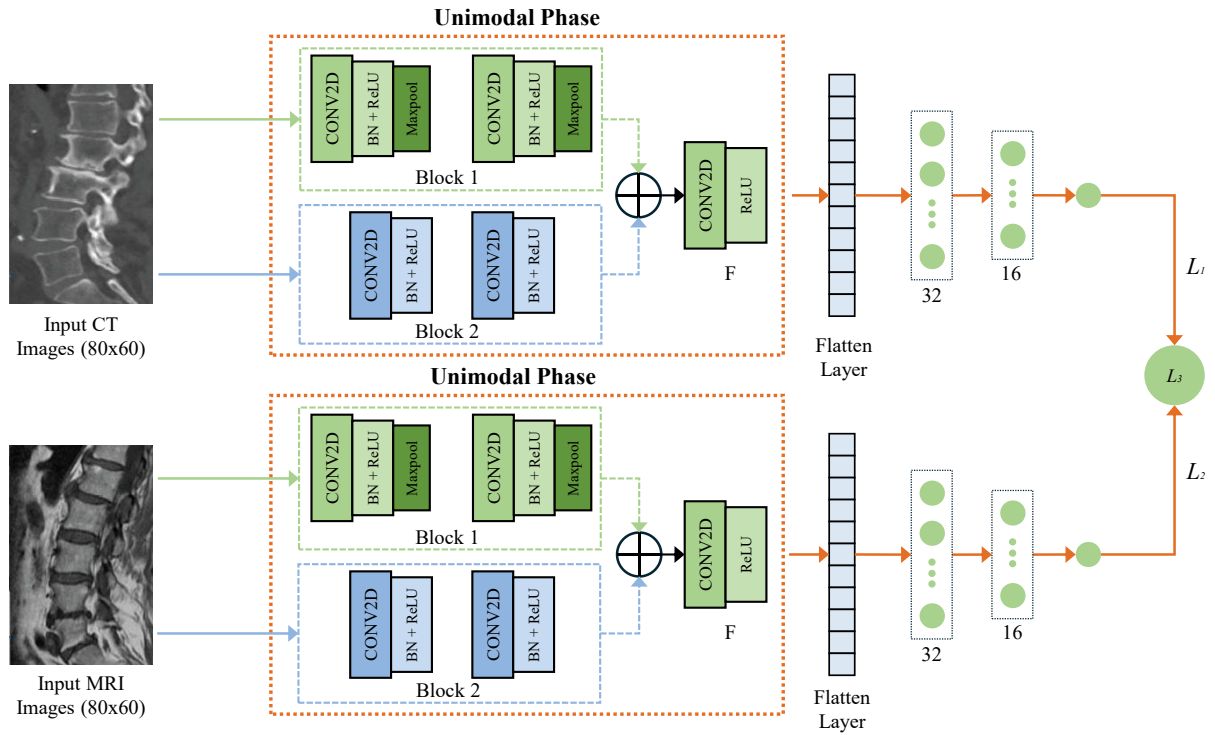

**Figure 13** Proposed multimodal architecture in [30]

In [69], the author proposed a DL-based framework for osteoporosis prediction from hip radiographs, aiming to address the limited DXA accessibility in Taiwan. The study included 139 hip X-rays from 134 elderly men and postmenopausal women, paired with BMD and T-score data. Four anatomical regions, the femoral neck, greater trochanter, Ward's triangle, and total hip were segmented using U-Net and U-Net++, both chosen for their encoder, decoder architecture and skip connections that preserve spatial resolution. Segmentation quality, measured by Intersection over Union (IoU), was satisfactory (≥0.5) across regions. The segmented and original images were then used for classification using VGG16, ResNet50, and DenseNet121, with DenseNet121 achieving the highest performance (74% accuracy, F1 = 0.71) for total hip classification without segmentation. Interestingly, segmentation-based classification degraded accuracy, likely because removing the surrounding anatomical context limited the CNNs' ability to capture global structural cues, such as cortical thickness continuity or pelvic alignment, which may be relevant to BMD estimation. Data augmentation through rotation, shifting, and scaling improved robustness, partially mitigating overfitting risks from the small dataset. A comparative analysis across architecture and preprocessing strategies is a strength, highlighting DenseNet121's advantage in reusing features across layers for fine-grained trabecular texture analysis. However, the study's small, imbalanced dataset, especially for the greater trochanter and Ward's triangle, restricted its statistical power and generalization. This work reinforced the feasibility of CNN-based hip X-ray analysis for opportunistic osteoporosis screening while also underscoring the importance of preserving anatomical context and expanding datasets for clinically reliable deployment.

A DL-based model, DeepSurv, was presented in [79] to predict future vertebral fractures from spine X-rays in a longitudinal cohort of 1,595 individuals aged 50–75 without baseline fractures. Using lateral lumbar spine radiographs and clinical data (age, sex, BMI, glucocorticoid use, and secondary osteoporosis), they trained a CNN-based survival model incorporating key point detection and vertebral patch extraction (L1–L5) for image preprocessing. The model was benchmarked against conventional methods such as the FRAX and Cox proportional hazard models. DeepSurv, particularly when using image data alone, achieved a higher C-index (0.614) than FRAX (0.547), highlighting its predictive advantage, even without clinical features. The preprocessing strategy of using non-masked L1–L5 vertebral patches yielded the best performance (AUROC 0.801). Although the model showed competitive results and introduced survival analysis into fracture prediction, its limitations included a small test set size, relatively low fracture event rates, and potential overfitting. Nevertheless, the approach demonstrated that opportunistically acquired spine X-rays analyzed via DL can aid in early fracture risk stratification, especially in settings where DXA is inaccessible.

Muhammad and Lee [25] developed a hybrid attention-driven DL framework for binary osteoporosis classification from knee radiographs, leveraging the KOP dataset of 372 labeled images (186 normal and 186 osteoporotic). To mitigate overfitting from the small dataset, extensive augmentation (zooming, cropping, and rotation) expanded the training set to 1,800 images. The architecture combined features from a pre-trained DenseNet-169, selected for its dense connectivity, which promotes feature reuse and gradient stability, with a custom attention module (AM) that incorporates both channel and spatial attention. Channel attention emphasized feature maps most correlated with osteoporotic patterns, such as trabecular rarefaction or cortical thinning, while spatial attention localized image regions most likely to exhibit bone deterioration, refining the model's focus on clinically relevant structures. The fused 1,792-dimensional feature vector was processed through a deeply tuned, fully connected network with progressive dimensionality reduction, dropout regularization, and batch normalization to enhance generalization. Compared with baselines such as ViT, ResNet50, and EfficientNetB0, the proposed model achieved a balanced trade-off between sensitivity (82.93%) and specificity (87.10%), with an F1-score of 0.8608 and the lowest loss (0.5834) among all tested architectures. While attention-enhanced fusion improved robustness and interpretability, its limitations included small dataset size, binary task design excluding osteopenia, and reliance on manual dataset curation. The study demonstrated that attention-augmented CNNs can extract discriminative, clinically aligned features, even from non-traditional imaging sites, such as the knee, offering a pathway for opportunistic osteoporosis detection in resource-limited screening scenarios.

Ref. [70] introduced a joint learning framework for multi-class osteoporosis classification from lumbar CT scans, aiming to provide a cost-effective alternative to DXA. The self-constructed dataset comprised 1,048 scans of L1 and L2 vertebrae, labeled as normal, osteopenia, or osteoporosis. The pipeline unified vertebral localization, segmentation, and classification in an end-to-end architecture (**Figure 14**), reducing the cumulative error from separate processing stages. A boundary heatmap regression branch improved segmentation precision by explicitly modeling vertebral edge likelihood, which is critical for accurate BMD-related feature extraction. The gated convolutional module adaptively modulates feature propagation, allowing the network to emphasize trabecular microarchitecture patterns while suppressing irrelevant background structures. Furthermore, the feature fusion module learned optimal weighting between L1 and L2 features, leveraging inter-vertebral consistency to enhance classification robustness. To counteract dataset imbalance, the authors implemented an instance- and class-based sampling strategy, ensuring that underrepresented categories contributed proportionally to training. The model achieved 93.3% classification accuracy and an AUC of 0.985, outperforming single-task and non-fusion baselines. While the multitasking design increased efficiency and reduced error propagation, the study's reliance on single-center data and the absence of external validation limited its generalizability. This work exemplified how integrated, multi-task learning with tailored architectural components can yield high-accuracy, clinically adaptable AI solutions for opportunistic osteoporosis screening using routine CT imaging.

Ref. [77] introduced MVCTNet, a multi-view DL framework for automated osteoporosis and osteopenia classification from sagittal CT scans, designed to bypass the limitations of prior approaches requiring manual ROI extraction. The dataset consisted of 2,883 patients labeled according to three categories (normal, osteopenia, and osteoporosis) using DXA-derived T-scores. Unlike single-view CNNs, MVCTNet generates two HU-clamped CT views that emphasize complementary anatomical and density cues, which are processed using separate ResNet-18 or EfficientNet-b0 feature extractors. A dissimilarity loss (**Figure 15**) enforces the learning of non-redundant representations, while a main task layer aggregates them for classification using both cross-entropy and ordinal regression losses to reflect disease progression. This combination not only improves class separability but also aligns the model's output with the ordinal nature of bone density decline. Ablation studies revealed that multi-view inputs outperform single-view baselines, and ordinal regression provides superior calibration over standard multi-class classification. Grad-CAM visualization further showed complementary attention patterns across views, supporting the interpretability of the approach. The strength of the method lies in its elimination of manual cropping, exploitation of domain-specific HU ranges, and explicit modeling of disease order. However, reliance on expert slice selection and the absence of 3D volumetric analysis may limit its scalability to fully automated pipelines. By integrating view diversity, representation dissimilarity, and ordinal loss, MVCTNet demonstrated why architecture-level innovations, beyond raw dataset size, are critical for robust, clinically relevant osteoporosis screening.

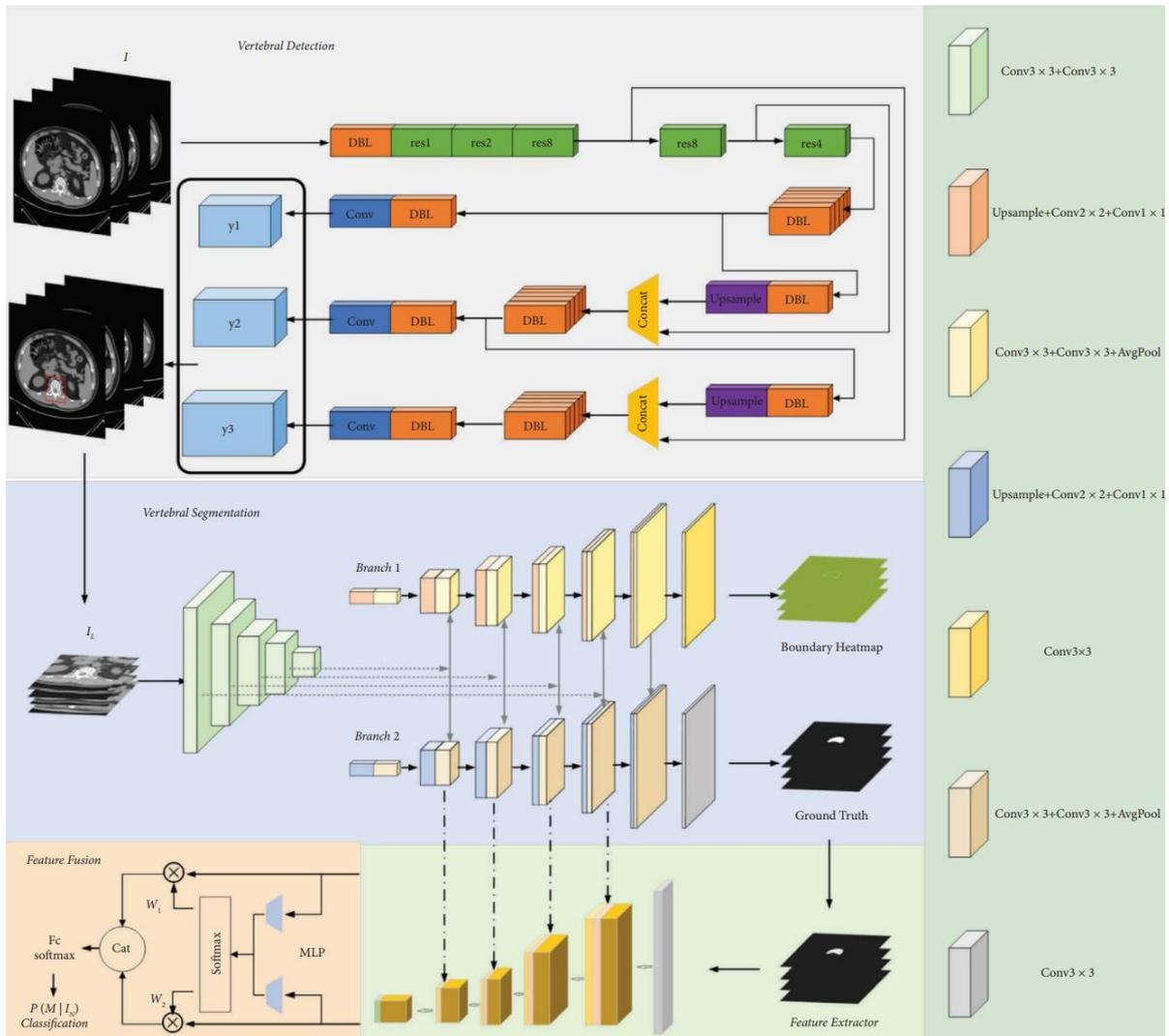

**Figure 14** Joint framework scheme proposed in [70]

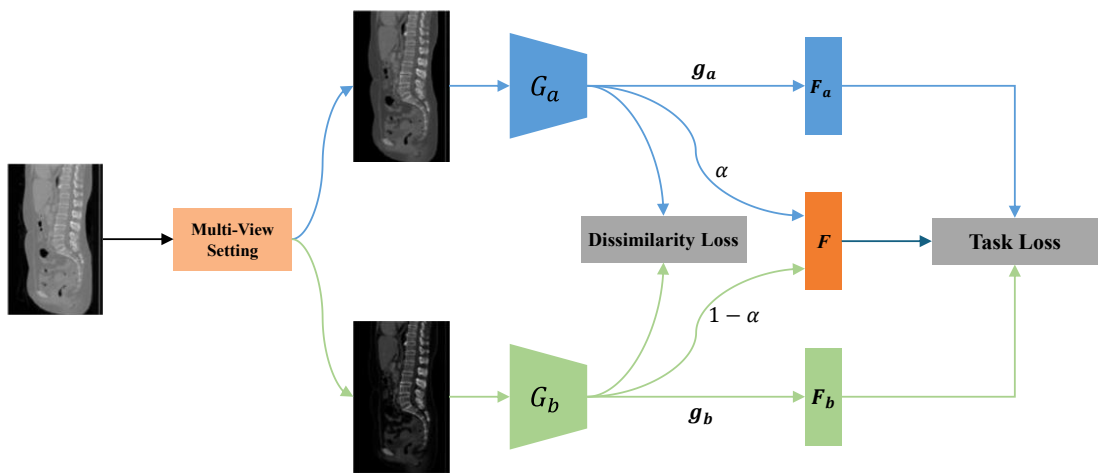

**Figure 15** Overview of the multi-view CT ensemble network proposed by Hwang et al. [77]

Kim et al. [28] addressed the clinical gap in short-term (<5 years) fracture risk prediction following an initial osteoporotic hip fracture by developing a CT-based DL model. Using digitally reconstructed radiographs (DRRs) derived from 3D hip CT scans of 1,480 patients (with 135 subsequent fractures), they trained a 2.5D ensemble model composed of three DenseNet branches (frontal, lateral, and axial views), combined through a decision aggregator. The task was to predict future osteoporotic fractures using survival-based outputs. The model was trained on 1,012 patients and tested on 468 temporally distinct cases. It achieved a C-index of 0.73 and AUCs of 0.74 over both 2 and 3 years, outperforming single-view image models and clinical models built with FRAX-equivalent variables. Notably, it also surpassed models based on 3D volumes and hip radiographs. The key strengths include its survival-aware design, use of multi-view DRRs (enabling richer feature representation), and validation on an independent temporal test set. However, limitations include reliance on a specific CT acquisition protocol, interpretability challenges despite Grad-CAM visualization, and generalizability concerns due to single-center data. The work provided a promising image-based solution for identifying high-risk patients requiring early intervention post-hip fracture.

In [81], Hu et al. developed a DL framework using CT radiography to predict subsequent osteoporotic VCF (OVCF) after an initial fracture. The study analyzed 23,383 CT images from 103 patients who experienced repeated OVCFs and 70 age-matched osteoporotic controls without fractures. The target vertebra (TV) was identified from scans acquired before the subsequent fracture, with images preprocessed to a uniform resolution. An Xception-based CNN was trained using the Adam optimizer, employing fivefold cross-validation on 80% of the patient data and independent testing on the remaining 20%. The model achieved an accuracy of 83.9% and ROC–AUC of 0.883 on the main test set, with stable performance on an independent cohort (81.7% accuracy). Class activation maps confirmed attention on fracture-prone vertebral regions, enhancing interpretability. Strengths include the use of real-world clinical imaging, lesion-focused interpretability, and consistent generalization across subgroups, whereas limitations include the single-center retrospective design, limited sample size, and lack of integration with clinical risk factors. Overall, the work demonstrated the feasibility of CT-based AI prediction for the early identification of high-risk patients, supporting proactive interventions to prevent refracture.

Despite these successes, CNN-based pipelines have practical limitations. Most are trained on small, single-center cohorts with DXA-derived labels, making them vulnerable to domain shift and label noise, resulting in frequent sensitivity drops for external cohorts. Many workflows still rely on manual cropping/ROI choices or post-hoc BMD formulas, reintroducing operator dependence and hindering scalability. Conversely, whole-image models can latch onto spurious contexts. The use of 2D radiographs to infer 3D microarchitecture makes osteopenia particularly challenging to detect, and segmentation can sometimes remove helpful global cues. Interpretability remains shallow (Grad-CAM ≠ causal explanation), while calibration, decision-curve analysis, and prospective validation are rarely reported, limiting clinical actionability. Finally, data/annotation demands, class imbalance, and population skew (in terms of age, sex, and ethnicity) raise concerns regarding fairness and generalizability, and many models are optimized for discrimination rather than time-to-event outcomes, which are necessary for fracture risk management.

### 5.3.3. Transformer-based Approaches

In recent years, Transformer architectures, originally developed for natural language processing, have gained significant traction in medical imaging owing to their ability to model long-range dependencies and contextual relationships more effectively than traditional CNNs. Unlike CNNs, which operate with local receptive fields, Transformers leverage self-attention mechanisms to capture global features across the entire input, making them particularly suitable for tasks requiring a holistic understanding of anatomical structures. This paradigm shift has inspired the development of vision Transformers (ViTs), hybrid CNN–Transformer models, and hierarchical Transformer architectures tailored for medical image analysis. In osteoporosis-related applications, Transformer-based models have shown promise in vertebral segmentation, bone quality assessment, and fracture risk prediction by integrating multi-view information, enhancing interpretability, and improving robustness against data imbalances. As the field evolves, Transformer-based approaches are expected to play a pivotal role in advancing the accuracy and generalizability of AI-powered osteoporosis diagnosis and prognosis systems.

Linyan Xue et al. [71] proposed FCoTNet (**Figure 16**), a Transformer-based model for classifying osteoporosis, osteopenia, and normal bone health from lumbar spine X-ray images, addressing limitations in traditional CNN-based methods that struggle with low-contrast and noisy radiographs. The study utilized a private dataset of 878 DXA-labeled anteroposterior and lateral lumbar X-rays from 439 patients and an external set of 38 images from 19 patients for robustness testing. FCoTNet enhances the Contextual Transformer Network (CoTNet) by incorporating multi-scale feature extraction with weighted fusion of different receptive fields, allowing for fine-grained texture extraction, which is crucial for osteoporosis diagnosis. The architecture integrates both convolutional and self-attention mechanisms to capture local and global dependencies. Evaluated via 5-fold cross-validation, the model achieved an accuracy of 78.29%, sensitivity of 69.72%, and specificity of 88.92%, outperforming baseline models, including CoTNet, ResNet, and ViT. The model also improved clinicians' diagnostic performance when used in assisted mode, with notable gains in inter-rater consistency. Although the model demonstrated strong performance internally, it exhibited reduced accuracy on the external dataset (55.26%), indicating limitations in generalization. Overall, FCoTNet exemplifies the potential of Transformer-style architectures in enhancing osteoporosis prediction from low-cost imaging modalities, such as X-ray.

Ref. [26] explored a Transformer-based DL approach for the binary classification of knee osteoporosis using X-ray images, addressing the limited exploration of non-traditional imaging sites for osteoporosis screening. The balanced dataset from Kaggle comprised 744 radiographs (372 normal and 372 osteoporotic), split into training, validation, and test sets. Images were pre-processed, resized, tensor-converted, and normalized to standardize the inputs. A Swin Transformer architecture, pre-trained on ImageNet, was fine-tuned within the PyTorch framework and optimized using Adam and Cross-Entropy loss. The model achieved an accuracy of 89.38 %, outperforming CNN baselines such as VGG16 (88%) and ResNet-50 (84%). Its strengths include a hierarchical attention mechanism for capturing both local and global features, robustness with limited data, and the use of a lightweight, reproducible pipeline (Figure XX). However, its limitations include reliance on a small, binary-labeled dataset without osteopenia representation and potential overfitting risk. This work demonstrated the feasibility of Transformer-based models for opportunistic osteoporosis detection from accessible imaging modalities, such as knee radiographs, offering a promising alternative in DXA-limited environments.

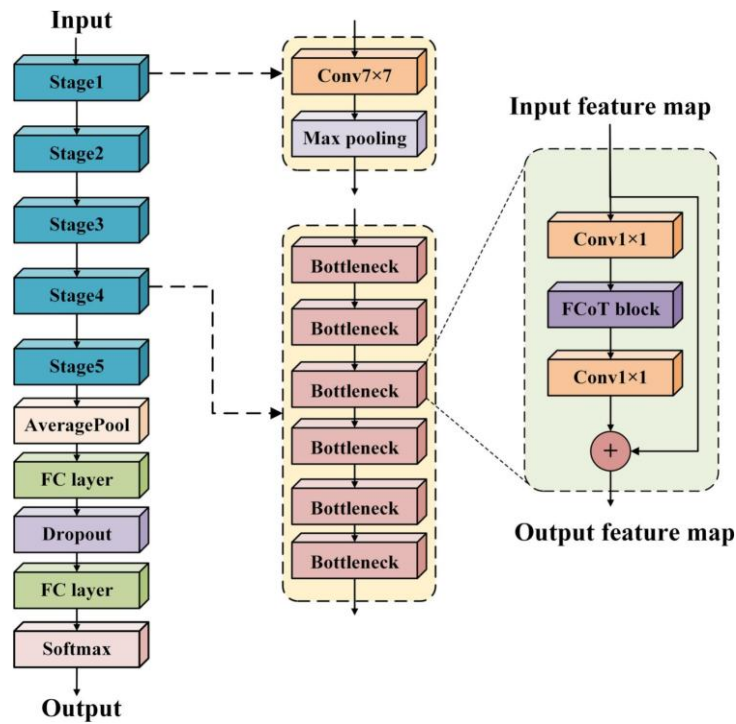

**Figure 16** Overall architecture of FCoTNet [71]

In [80], a Transformer-based DL model was introduced for predicting post-radiotherapy osteoporosis (OP) in cervical cancer patients using baseline pelvic CT scans, with the aim of personalizing treatment and mitigating radiation-induced bone damage. The study retrospectively analyzed CT data from 269 patients across three centers (2021–2022), incorporating both imaging and clinical data, such as age and dosimetry parameters (e.g., pelvic V20/V40 and LSS-V10). The model leveraged a 2.5D data preparation strategy, extracting seven CT slices per patient around the ROI, and employed a multi-head self-attention transformer to fuse features (see **Figure 17**). Additional enhancements include multi-instance learning (MIL) using Bag-of-Words (BoW), histogram features, and ensemble fusion methods. Evaluation on both internal and external validation sets revealed strong predictive performance, with the combined Transformer–clinical model achieving AUCs of 0.913 (training set), 0.900 (internal validation set), and 0.977/0.964 (external validation sets). This model surpassed other configurations, including CNNs, MIL, and ensemble methods. Its strength lies in the integration of multi-slice features and robustness across datasets, while its limitations include the use of HU instead of DXA for OP diagnosis and relatively small external cohorts. The work not only provides a scalable pipeline for radiotherapy-induced OP prediction but also promotes bone-sparing radiotherapy strategies, such as Pelvic Bone Marrow-Sparing Intensity-Modulated Radiotherapy (PBMS-IMRT), and potential clinical transfer to other pelvic malignancies.

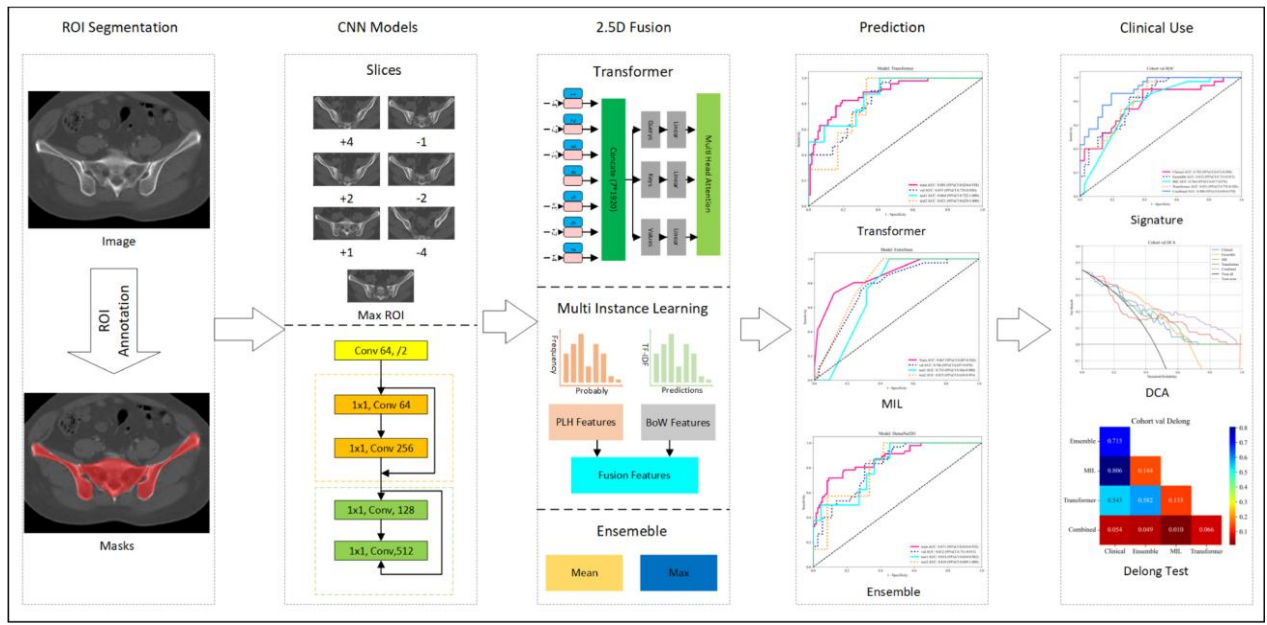

**Figure 17** Overall study flowchart proposed by Huang et al. [80]

Ref. [78] presented a novel framework for the automatic identification of VCF types (A, B, and C) in CT images spanning the C3–L5 regions, addressing the diagnostic challenges posed by inter-observer variability and limited labeled data. The dataset consisted of 2,820 annotated CT images collected from a tertiary care hospital, augmented using the Extended Deep Convolutional Generative Adversarial Network (DCGAN) [85] and Progressive Growing of GANs (PGGAN) [86] techniques to combat data imbalance and enhance model generalization. The study introduced a weighted ensemble model integrating ViT [87] with CNN-based architectures (VGG16 and ResNet50), leveraging ViT's global self-attention for long-range dependency modeling and CNNs' strength in local feature extraction. The ensemble was trained using a weighted averaging mechanism based on model performance and optimized using Adam [88] with regularization techniques. The evaluation results showed that ViT alone achieved 86.28% accuracy with traditional augmentation, which improved to 91.28% with PGGAN, and the final ensemble model achieved 93.68% accuracy, outperforming all baselines. Its strengths include effective hybrid modeling, robust augmentation, and superior performance in multi-class VCF classification. However, this method relied on GAN-generated images and lacked external validation. The study demonstrated the transformative role of ViT–CNN ensembles in fine-grained fracture classification and suggests their broader clinical utility in orthopedic diagnostics.

Despite their promise, Transformer-based pipelines remain data-hungry and fragile under a domain shift, as many studies utilize small, single-center cohorts and surrogate labels (e.g., HU- or RT-induced OP instead of DXA/fractures), and exhibit sharp performance drops on external datasets. Patch/2.5D tokenization and heavy down sampling can miss fine trabecular cues and true 3D anisotropy, while full-3D Transformers are computation- and memory-intensive, pushing teams towards compromises that blunt sensitivity, especially for osteopenia. Hybrid designs and GAN-augmented training may inflate results and introduce synthetic artifacts. Attention maps improve plausibility but do not guarantee causal explanations. Class imbalance, population skew, and limited multi-ethnic validation raise concerns about fairness, and few works have reported calibration, decision-curve analysis, or prospective testing, which are necessary for clinical use. Finally, complex multi-view/ensemble stacks increase engineering burden and deployment latency, hindering reproducibility and real-world integration.

*5.3.4. Self-Supervised Learning*

Self-supervised learning (SSL) has emerged as a powerful paradigm in medical imaging, particularly in domains like osteoporosis detection, where labeled data are scarce, expensive, and often subject to inter-observer variability. Unlike supervised methods that rely heavily on annotated datasets, SSL leverages inherent structures within the data to generate surrogate labels and pretext tasks, such as image inpainting, jigsaw solving, context prediction, and contrastive representation learning. This allows models to learn meaningful feature representations from large volumes of unlabeled images. These learned representations can then be fine-tuned for downstream tasks such as fracture detection, bone quality assessment, or risk prediction with minimal labeled samples. Recent SSL frameworks, such as SimCLR, MoCo, and masked autoencoders, have demonstrated strong generalizability across imaging modalities (e.g., X-ray, CT, and DXA) and tasks. In osteoporosis-related applications, SSL has shown promise in enhancing model robustness, reducing annotation costs, and enabling efficient transfer learning across datasets or institutions. As the field advances, SSL is expected to play a pivotal role in democratizing AI-driven osteoporosis screening by effectively utilizing unlabeled clinical data.

Ref. [76] introduced a novel semi-supervised learning framework for osteoporosis diagnosis using dental panoramic radiographs, leveraging synthetic data generated by Denoising Diffusion Probabilistic Models (DDPM) and a sinusoidal threshold decay mechanism (**Figure 18**). The study addressed the challenge of limited labeled medical data by generating 45,000 realistic synthetic images from 12,000 unlabeled panoramic X-rays, validated by dental experts for quality. Using a WideResNet backbone, the model was trained on 399 labeled images and synthetic unlabeled data, avoiding complex preprocessing to maintain clinical feasibility. A key methodological innovation is the sinusoidal threshold decay for consistency loss regularization, which dynamically adjusts pseudo-labeling thresholds to enhance convergence and robustness. Comparative experiments showed that the proposed model outperformed traditional semi-supervised approaches, even when trained on real unlabeled data, achieving a maximum accuracy of 80.10%. Its strengths include improved generalization using synthetic data, enhanced threshold scheduling, and reduced reliance on large, annotated datasets. Its limitations include the single-institution dataset and reliance on handcrafted thresholds. This study demonstrated the potential of self-supervised and synthetic data-driven techniques in scalable, cost-effective osteoporosis screening.

Hwang et al. [75] proposed a segmentation-for-classification framework enhanced with SSL to predict osteoporosis from hand and wrist X-ray images, offering an affordable, DXA-free screening alternative. The dataset included 192 labeled images paired with DXA-derived T-scores and 1,154 unlabeled images used for pretraining. Their pipeline begins with a probabilistic U-Net architecture for uncertainty-aware segmentation of the ulna, radius, and metacarpals, framed as an optimal transport problem for aligning predictions with multi-annotated masks. From these masks, they generate bone-specific image segments and apply an extended multi-crop augmentation strategy tailored to radiographic sparsity. The pretext SSL phase explores SimCLR, SupCon, SwAV, and VICReg on these augmented segments to learn robust features without labels. The final classification is achieved using a linear head atop the frozen encoder, aggregating predictions across all segments per subject. Among all SSL methods, SimCLR achieved the best results (F1-score 0.68, AUC 0.85). The model also identified the 2nd metacarpal, radius, and ulna as the most predictive. Its strengths include improved interpretability, reduced label dependency, and strong generalization with limited data; however, the lack of external validation and dependency on manual segmentation annotations via SAM are limitations. This study highlighted the potential of SSL for osteoporosis screening using peripheral radiographs in data-constrained environments.

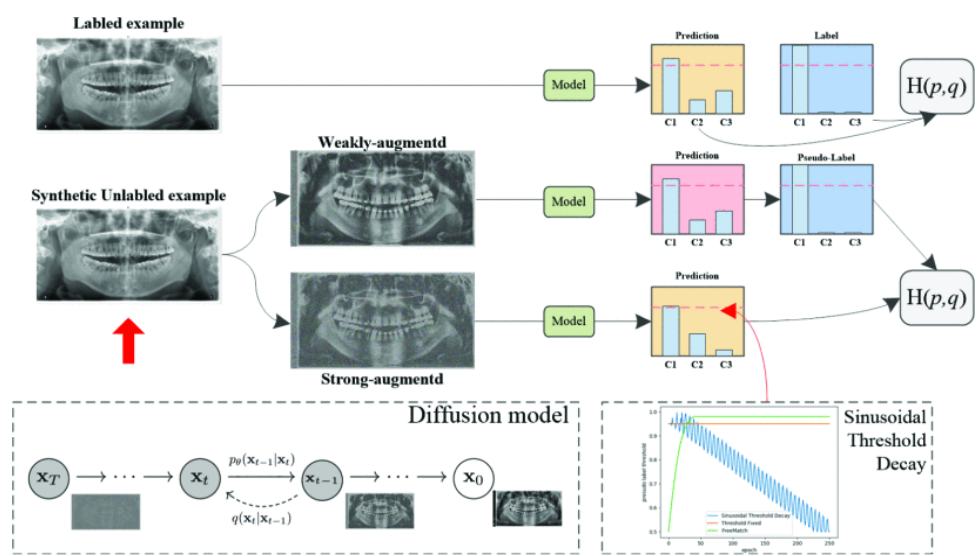

**Figure 18** Overall architecture proposed in [76]

Ref. [74] proposed a few-shot learning (FSL) framework to address the challenge of diagnosing osteopenia and osteoporosis using knee X-ray images in data-scarce environments. The study utilized a novel two-stage pipeline in which deep CNNs (VGG16, ResNet50, and Xception), pre-trained on ImageNet, are fine-tuned using a chest X-ray dataset (pneumonia vs. normal) to extract radiographic features. The core task uses support and query sets of knee X-rays across two cohorts, Cohort-1 (189 images) and Cohort-2 (609 images), with diagnostic labels based on DXA-derived T-scores. Feature embeddings of query images are compared to support sets via Euclidean distance, and the resulting vectors are passed to AutoML classifiers (e.g., XGBoost) for the final prediction. Evaluation across multiple rounds showed that the FSL models outperformed both junior and senior radiologists in terms of accuracy (up to 74.5%) and sensitivity, although sometimes at the expense of specificity. Grad-CAM visualizations further enhance interpretability by highlighting discriminative knee regions. While the model generalizes well across cohorts and offers a practical AI-assisted diagnostic tool in low-data settings, its reliance on a specific imaging modality and lack of multimodal fusion are notable limitations. The study demonstrated the viability of FSL for osteoporosis-related applications and supported its integration into diagnostic workflows.

Siddiqa et al. [27] proposed a computer-aided diagnosis system for knee osteoporosis using transfer learning and stacked DL modules to address the limitations of traditional manual and handcrafted feature-based approaches. The study utilized knee X-ray images sourced from three publicly available datasets: OKX Kaggle Binary, OKX Kaggle Multiclass, and KXO-Mendeley Multiclass, encompassing binary and multi-class labels for osteoporosis detection. The main objective was to accurately classify knee conditions (normal, osteopenia, and osteoporosis) using DL to enable early and efficient diagnosis. The methodology integrates transfer learning via a pre-trained ResNet50 model for initial feature extraction, followed by five sequential Conv-ReLU-MaxPooling blocks that progressively enhance features, culminating in a dense layer-based classification module. Evaluations showed impressive performance across all datasets, achieving up to 98.4% accuracy and consistently high precision, recall, and F1-scores, indicating strong diagnostic potential. The strengths of this approach lie in its robust generalization, ability to handle multi-class classification, and minimal reliance on large, annotated datasets owing to transfer learning. However, the model's computational complexity and interpretability may limit real-time deployment. Overall, the study demonstrated that combining transfer learning with deep hierarchical feature enhancement can significantly improve the accuracy and clinical utility of osteoporosis diagnosis from knee X-rays.

Ref. [29] addressed the challenge of fracture risk classification in older adults using DXA scans, a modality widely used for BMD assessment but rarely leveraged for image-based risk prediction. The study involved a dataset of 1488 DXA images (hip, spine, forearm, and whole body) from 526 subjects (478 fallers and 48 non-fallers), categorized into low, moderate, and high fracture risk groups based on past fracture history. The primary objective was to predict fracture risk using DL models trained on DXA images with and without accompanying clinical data. The authors utilized both CNNs (VGG-16 and ResNet-50) and Transformer-based models (ViT-S), with feature extraction performed via self-supervised learning techniques: Momentum Contrast (MoCo) and DINO. These features are fed into an MLP classifier and, in some configurations, fused with clinical variables such as vitamin D, calcium levels, ALM/BMI, and height. The evaluation metrics included F1-micro score and ROC-AUC, with the best model (ViT-S + DINO + clinical data) achieving a ROC-AUC of 75.1% and F1-micro of 63.6%, outperforming baseline CNN models. While the strengths include the novel use of ViT with self-supervised pretraining and multimodal data integration, the limitations include a small sample size, class imbalance, and low image resolution. The study underscored the potential of Transformer-based models for clinical risk stratification from DXA images and opens avenues for enhancing predictive accuracy using higher-quality datasets and explainable AI tools.

Building on these gains, real-world deployment exposes limitations: SSL often assumes abundant, in-distribution unlabeled scans and hinges on augmentation choices, and domain shifts can break contrastive or masking objectives and erase subtle trabecular cues. Synthetic data and pseudo-labels may propagate artifacts and cohort bias, inflating internal scores but collapsing on external, multi-center tests. Common 2D slice/patch pretraining (and MAE's low-frequency bias) misses true 3D anisotropy, making osteopenia difficult to detect, while dependencies on manual ROIs/segmentations or brittinflatesigns reduce scalability and reproducibility. Finally, calibration, decision-curve analysis, and prospective validation are rarely reported, fairness issues persist with class imbalance and population skew, and computation-heavy pretraining limits used in low-resource settings.

### 5.3.5. Explainable AI

As AI systems gain traction in medical imaging for osteoporosis detection, the need for interpretability and transparency becomes increasingly critical. While DL models, such as CNNs and Transformers, offer impressive diagnostic accuracy, their "black box" nature often hinders clinical trust and adoption. Explainable AI (XAI) methods aim to bridge this gap by providing visual or quantitative insights into model decisions, thereby enhancing model accountability, enabling error analysis, and facilitating physician-AI collaboration. In the context of osteoporosis, where subtle structural changes in bone tissue can be clinically significant, XAI techniques such as saliency maps, Grad-CAM, LIME, and SHAP have been employed to highlight image regions or clinical features that are most influential to the predictions of AI. This section explores how XAI has been integrated into osteoporosis-related studies, evaluates its effectiveness in improving model interpretability, and discusses its role in advancing the clinical applicability of AI-assisted diagnostic systems.

Huang et al. [72] proposed a robust DL framework for opportunistic osteoporosis diagnosis using CT imaging, addressing the key limitations of current AI models, such as limited labeled data, device-specific DXA bias, and neglect of anatomical priors. The study leveraged a large dataset comprising over 3,000 labeled and 30,000 unlabeled vertebral CT instances across multiple hospitals. Their objective was to perform accurate osteoporosis classification, BMD regression, and vertebra localization using a unified, generalizable model. The methodology integrates a texture-preserving self-supervised learning (TP-SwAV) module to exploit unlabeled CT scans, a Mixture of Experts (MoE) decoder for handling DXA device variability, and a multi-task learning strategy to jointly supervise diagnosis, regression, and spatial localization. Evaluations across four clinical sites (including an external test site) showed that their model, particularly the ResNet34+MoE variant with unified loss, achieved superior performance, with F1-scores exceeding 86% in internal validation and 83% in external tests. The strengths of this work include high generalizability, robustness across imaging devices, and improved diagnostic interpretability via integrated supervision, while the limitations include reliance on CT (which may not always be available) and model complexity. Overall, the study demonstrated that combining texture-aware self-supervision, expert-guided decoding, and clinically informed task integration significantly enhances the scalability and accuracy of AI-driven osteoporosis screening.

Currently, XAI for osteoporosis imaging remains uncommon, as most efforts employ post-hoc methods (saliency/Grad-CAM, LIME/SHAP), whose attributes are sensitive to preprocessing and scanner/vendor shifts and can emphasize non-causal context. Cohorts are small and modality-specific, standardized uncertainty-aware metrics are rare, and multi-center/prospective evidence is limited; therefore, improvements in calibration, reader performance, and clinical decision-making impact are not yet established.

### 5.4. Clinical Readiness Gaps

Despite notable progress in AI-powered osteoporosis detection, the current literature reveals recurring structural and methodological issues that constrain its robustness, comparability, and readiness for clinical deployment. The synthesis of these limitations is summarized below.

**Data-related limitations:** A substantial proportion of studies relied on datasets from single institutions or geographically limited regions, resulting in a narrow representation of patient demographics, imaging equipment, and acquisition protocols. Such homogeneity increases the risk of overfitting to site-specific characteristics, limiting generalizability across diverse clinical environments. Approaches such as federated learning can expand multi-site collaboration without centralizing data, helping mitigate these constraints [89].

Sample sizes are frequently modest, particularly for DXA and MRI, which constrains the statistical power to detect subtle patterns, hinders the training of large-scale deep neural networks, and increases the likelihood of model instability. Publicly available, well-annotated datasets remain scarce, with most accessible repositories focusing on X-ray images. This lack of open, multimodal datasets prevents consistent cross-study benchmarking and slows methodological innovation.

External validation, when performed, often results in a marked drop in performance compared to internal testing. This degradation is typically attributable to domain shift, differences in scanner models, reconstruction algorithms, imaging protocols, and patient populations, which existing studies have rarely addressed systematically. Although domain adaptation and harmonization techniques [90] are well-established in other areas of medical imaging, their adoption in osteoporosis-related AI research remains limited.

**Scope imbalance:** The current body of work is heavily skewed towards X-ray and CT modalities, which, while valuable, do not capture the full clinical spectrum of bone health assessment. DXA, the clinical gold standard, and MRI, which are uniquely suited for marrow composition and microarchitecture analysis, remain underrepresented in AI research despite their clear diagnostic relevance.

From a task perspective, most studies have focused on osteoporosis classification, often framed as a binary or three-class problem based on BMD thresholds. While this approach aligns with WHO diagnostic criteria, it fails to leverage the full potential of AI for more complex and clinically impactful applications, such as fracture risk prediction, longitudinal disease monitoring, and opportunistic screening from non-dedicated scans. The underrepresentation of these tasks limits the potential of AI to contribute to preventive care, personalized treatment planning, and early intervention strategies.

**Methodology and reproducibility issues:** Methodological diversity across studies, while indicative of active exploration, poses significant challenges for reproducibility and comparative evaluation. Many models depend on single-view or cropped ROIs, which may exclude critical anatomical context. In contrast, multi-view, 3D volumetric, and context-aware approaches are rare, even though they can more accurately capture bone morphology and biomechanical relationships.

Evaluation protocols also vary widely. While accuracy, AUC, sensitivity, and specificity are commonly reported, few studies provide calibration analyses, decision-curve evaluations, or uncertainty quantification, which are metrics that directly inform clinical reliability and utility. Moreover, differences in preprocessing pipelines, annotation guidelines, and dataset splits make direct comparison of results difficult and potentially misleading.

Reproducibility is further hampered by the limited release of source code, trained model weights, and annotation protocols. Community reporting standards, such as CLAIM for imaging AI [91] and CONSORT-AI [92]/SPIRIT-AI [93] for trials, provide concrete checklists to improve transparency and comparability, yet they are inconsistently applied. Without these resources, independent verification and reimplementation become impractical, slowing the translation of research findings to deployable clinical tools.

**Clinical translation challenges:** Technical performance in retrospective datasets does not guarantee clinical impact. However, only a small subset of models has undergone prospective evaluation or real-world deployment trials. Integration into clinical infrastructure, such as PACS/RIS systems or radiologist-in-the-loop workflows, is rare, leaving key questions regarding scalability, user acceptance, and workflow compatibility unanswered.

XAI methods, although occasionally used, often remain limited to visual attention maps (e.g., Grad-CAM) without providing interpretable, clinician-oriented rationales that could support trust and adoption. Additionally, few studies explicitly address the regulatory and operational requirements for deployment, including interoperability with standards such as HL7 and FHIR, compliance with data privacy regulations, and strategies for safe system behavior under uncertain inputs.

Without targeted efforts to bridge this gap, from regulatory approval to seamless workflow integration, the practical utility of AI for osteoporosis detection will remain largely unrealized, regardless of the reported retrospective performance.

In summary, the literature demonstrates rapid technical progress but remains fragmented in scope, methodology, and translational readiness. Addressing the intertwined challenges of dataset diversity, balanced task and modality coverage, methodological standardization, and clinical integration will be essential for transitioning from promising prototypes to robust, scalable, and trustworthy AI systems for osteoporosis care. These considerations directly inform the recommendations and future directions outlined in the next section.

## 6. Emerging Trends and Future Directions

### 6.1. Advances in AI Methodologies and Multi-modal Integration

The trajectory of AI for osteoporosis detection and fracture risk assessment is moving from single-modality, CNN-centric pipelines towards representation-rich and data-efficient paradigms. ViT and hybrid CNN-Transformer models improve global context modeling in spine, hip, and knee imaging, helping capture diffuse trabecular changes and cortical thinning that span large receptive fields. Self-supervised pretraining (contrastive, masked-image modeling, and teacher–student distillation) is increasingly being used to leverage large volumes of unlabeled DXA, X-ray, and CT data, thereby reducing the dependence on scarce annotations and improving transfer across scanners and patient populations.

Parallel progress has been achieved in multimodal fusion studies. Architectures that combine images with clinical variables (age, sex, BMI, prior fracture, and medications) or fuse across modalities (DXA + X-ray, CT + MRI when available) yield more holistic phenotypes of bone quality. Late-fusion and cross-attention designs allow models to weigh modality-specific evidence, while tabular-image co-training improves calibration at clinically relevant thresholds. Generative models (diffusion/GANs) are being explored for data augmentation, domain translation (e.g., CT kernel harmonization), and simulation of rare phenotypes, although they require rigorous safeguards against bias amplification and leakage. Finally, opportunistic imaging, which involves mining routine CTs, or dental radiographs acquired for other purposes, continues to grow as a practical route to large, real-world datasets and low-friction clinical impact.

*6.2.Underexplored Areas and Strategic Recommendations*

The current literature is heavily concentrated on classification tasks, with limited attention paid to longitudinal modeling, subtle fracture detection, and real-world deployment considerations. Moreover, data limitations, inconsistent evaluation practices, and underdeveloped governance frameworks continue to restrict clinical readiness. Addressing these issues requires targeted research efforts, standardized resources, and robust validation strategies.

***Underexplored problems:***
- Longitudinal modeling of disease progression and time-to-fracture prediction using serial imaging and clinical follow-up.
- Fracture detection beyond vertebrae, including subtle hip and wrist injuries, and integration into opportunistic screening pathways.
- Robust handling of label noise and weak supervision, such as variability in radiology reports, T-scores, and automated vertebral labeling.
- Equity and domain generalization analyses across scanners, vendors, and healthcare settings, including low-resource environments.
- Real-world economic evaluation, covering cost-effectiveness, workflow impact, and outcomes from prospective, in-workflow trials.

***Method and dataset roadmap:***
- Establish open, multi-center benchmark datasets for DXA, X-ray, CT, and MRI, with standardized geographic, device, and temporal splits, including longitudinal cohorts.
- Adopt pre-registration and transparent reporting standards (e.g., TRIPOD-AI and CONSORT-AI), with code, model cards, and data sheets where feasible.
- Expand evaluation metrics beyond discrimination to include calibration, decision-curve analysis, reclassification indices, and for risk models, time-dependent C-index.
- Conduct rigorous external validation and stress testing, covering subgroups, image quality, out-of-distribution detection, and acquisition variability.
- Prioritize uncertainty quantification and design human–AI interaction studies to define safe operational thresholds.

***Modeling recommendations:***
- Leverage self-supervised and weakly supervised learning to utilize large unlabeled archives, coupled with active learning for efficient expert annotation.
- Use hybrid CNN–Transformer architectures for both fine-grained texture and global morphology analysis; explore graph- or region-centric reasoning for vertebra-level detail.

- Integrate explainability by design, including concept bottlenecks, anatomically aligned attention maps, counterfactuals, and structured rationales.
- Apply privacy-preserving and federated learning for multi-site collaboration, paired with harmonization and domain adaptation to mitigate scanner/protocol shifts.
- Ensure model calibration for intended clinical roles (screening vs. confirmatory) and implement fail-safe protocols for uncertain cases.

*Clinical translation and governance:*
- Design prospective, in-workflow trials with radiologist involvement, clear PACS/EHR integration, and measurement of real-world outcomes (e.g., reading time and fracture prevention).
- Establish multidisciplinary governance boards (radiology, endocrinology, data science, ethics, and legal) to oversee bias audits, model updates, and incident responses.
- Maintain comprehensive documentation of intended use, contraindications, and monitoring plans, supported by MLOps for version control, auditability, and post-market surveillance.

In summary, the field is trending towards data-efficient representation learning, multimodal fusion, and clinically aware evaluation. Converting these advances into reliable, equitable, and regulatory tools requires stronger datasets and benchmarks, rigorous external validation and calibration, actionable explainability, and careful integration into real clinical workflows.

## 7. Conclusion

AI has emerged as a transformative force in the domains of osteoporosis detection and fracture risk assessment, offering new possibilities for early diagnosis, personalized risk stratification, and improved patient outcomes. This survey systematically reviewed state-of-the-art AI approaches across various imaging modalities, including DXA, X-ray, CT, and MRI, and clinical tasks, such as osteoporosis classification, fracture detection, and risk prediction. We categorized and compared methods based on AI paradigms, including classical ML, CNNs, Transformer-based models, self-supervised learning, and XAI techniques.

Our analysis highlights the growing methodological sophistication in this field, alongside persistent challenges, such as limited data diversity, lack of external validation, and the need for greater model interpretability. Emerging trends point towards the adoption of Transformer architecture, multimodal fusion, and self-supervised learning, with increasing emphasis on clinical validation and real-world deployment. By bridging the gap between algorithmic innovation and clinical utility, AI-based systems have the potential to redefine how osteoporosis is detected, monitored, and managed in the years to come.

**Acknowledgment**

This work was supported by the National Research Foundation of Korea (NRF) grant funded by the Korea government (MSIT) (No. RS-2023-00217471).